\documentclass[twoside,leqno]{article}

\usepackage{siamproceedings}

\usepackage[T1]{fontenc}
\usepackage{amsfonts}
\usepackage{graphicx}
\usepackage{epstopdf}
\usepackage{enumitem}
\usepackage{algorithmic}
\ifpdf
  \DeclareGraphicsExtensions{.eps,.pdf,.png,.jpg}
\else
  \DeclareGraphicsExtensions{.eps}
\fi

\newsiamremark{remark}{Remark}
\newsiamremark{hypothesis}{Hypothesis}
\crefname{hypothesis}{Hypothesis}{Hypotheses}
\newsiamthm{claim}{Claim}
\newsiamremark{conjecture}{Conjecture}
\newsiamthm{question}{Question}
\newsiamthm{problem}{Problem}
\newcommand\E{\mathbb{E}}
\newcommand\eps{\varepsilon}
\newcommand\card[1]{\left| {#1} \right|}
\newcommand\inner[2]{\langle{#1},{#2}\rangle}
\newcommand\norm[1]{\| #1 \|}
\newcommand\inst{\mathcal{I}}
\newcommand\val{{\sf val}}
\newcommand\supp{{\sf supp}}

\usepackage{amsopn}

\begin{document}

\title{\Large The Lens of Abelian Embeddings}
    \author{Dor Minzer}

\date{\vspace{-5ex}}

\maketitle






\begin{abstract} 
We discuss a recent line of research investigating inverse theorems with respect to 
general $k$-wise correlations, and explain how such correlations arise in different contexts in mathematics. We outline some of the
results that were established and their applications 
in discrete mathematics and theoretical computer science. We also mention some open problems
for future research.
\end{abstract}

\section{Introduction.}\label{sec:intro}
We begin the discussion 
about the study of $k$-wise correlations with motivating examples.
\subsection{Arithmetic Progressions.}\label{sec:AP}
Our first example is the problem 
of upper bounding the size of sets 
of vectors with no \emph{$k$-term arithmetic progressions}.
\begin{definition}
A $k$-term arithmetic progression in $\mathbb{F}_p^n$ is a 
progression of the form  
$x,x+a,x+2a,\ldots,x+(k-1)a$, where $x\in\mathbb{F}_p^n$ and 
$a\in\mathbb{F}_p^n\setminus\{\vec{0}\}$. 
We say a set $A\subset\mathbb{F}_p^n$ is a $k$-AP free set if it 
contains no $k$-term arithmetic progressions.
\end{definition}

\subsubsection{Arithmetic Progressions of Length $3$.}\label{sec:intro_AP}
Using a density-increment approach, Meshulam~\cite{Meshulam} (following an argument of Roth~\cite{Roth}) used Fourier-analytic tools 
to give an upper bound on the size of a set $A\subseteq\mathbb{F}_p^n$
that contains no $3$-term arithmetic progressions, where $p$ is a prime.
Towards this end Meshulam expresses the condition that $A\subseteq\mathbb{F}_p^n$ contains no $3$-term arithmetic progressions analytically as 
$\E_{x,a\in\mathbb{F}_p^n}[1_A(x)1_A(x+a)1_A(x+2a)] = \mu(A)p^{-n}$, where $\mu(A) = |A|/p^n$ is the relative size of $A$. Taking the normalized indicator $f_A(x) = 1_A(x) - \mu(A)$ one concludes that if $\mu(A) \geq 2p^{-n/2}$, then
\begin{equation}\label{eq:3AP}
\E_{x,a\in\mathbb{F}_p^n}[f_A(x)f_A(x+a)f_A(x+2a)] = -\mu(A)^3+\mu(A)p^{-n}
\leq -\frac{1}{2}\mu(A)^3,
\end{equation}
and in particular the left hand side has noticeable absolute value. 
What can be said about the structure of the function $f_A$ in that case, namely in the case that its values on the points $x,x+a,x+2a$
has non-trivial correlation? 

To answer this question we use the Fourier expansion of $f_A$, which allows us to write 
$f_A(x) = \sum\limits_{\alpha\in\mathbb{F}_p^n}\widehat{f}_A(\alpha)e^{2\pi{\bf i}x\cdot \alpha}$, 
where $x\cdot \alpha = \sum\limits_{i=1}^{n}x_i\alpha_i$ and  
$\widehat{f}_A(\alpha) = \E_{x\in\mathbb{F}_p}[f_A(x)e^{-2\pi{\bf i}x\cdot \alpha}]$. 
An elementary calculation shows that
\[
\E_{x,a\in\mathbb{F}_p^n}[f_A(x)f_A(x+a)f_A(x+2a)]
=
\sum\limits_{\alpha\in\mathbb{F}_p^n}\widehat{f}_A(\alpha)^2\widehat{f}_A(-2\alpha),
\]
which quickly gives that there exists $0\neq \alpha\in\mathbb{F}_p^n$ such that $\card{\widehat{f}_A(\alpha)}\geq \Omega_p(\mu(A)^2)$. In words, $f_A$ must have a noticeable correlation with some Fourier character.

From this information one can deduce that there exists an affine subspace $W\subseteq\mathbb{F}_p^n$ of 
codimension $1$ such that
$\frac{\card{A\cap W}}{\card{W}}\geq \mu(A) + \Omega_p(\mu(A)^2)$. In words, the density of $A$ inside $W$ is significantly larger than the overall density of $A$.
By iterating this argument, Meshulam eventually gets to a subspace $W'$ of codimension $O_p(1/\mu(A))$ in which the density of $A$ is close to $1$. In that case $A$ must clearly contain a $3$-term arithmetic progression.

While the quantitative bound obtained by Meshulam, which is $\mu(A)\leq O_p\left(\frac{1}{n}\right)$, is by now obsolete thanks to the polynomial method~\cite{CrootLevPach,EllenbergGijswijt}, the Fourier-analytic approach underlying the argument plays a pivotal role in many subsequent results in additive combinatorics, as discussed next.
\subsubsection{Longer Progressions.}
The finite field version of Szemer\'edi's theorem~\cite{Szemeredi} asserts that 
for $p>k$, for any $\delta>0$ and sufficiently large $n\geq n_0(\delta,p)$, a set $A\subseteq \mathbb{F}_p^n$ with $\mu(A)\geq \delta$ must contain a $k$-term arithmetic progression. In other words, the measure of a set $A\subseteq \mathbb{F}_p^n$ with no $k$-term arithmetic progressions is a vanishing function of $n$. This result is a
significant extension of Meshulam's result, and in its proof Szemer\'edi introduces several very impactful ideas such as the Szemer\'edi's regularity lemma. 
Overall, Szemer\'edi's argument 
is much more combinatorial in nature and has no resemblance to Meshulam's argument at all.
Additionally, while the ideas used by Szemer\'edi turn out to be useful in a much broader context, they often have significant drawbacks, particularly with regard to the quantitative bounds they give. For example, the bound on $n_0(\delta,p)$ coming from Szemer\'edi's argument is tower-type.

Motivated by finding a higher order analog of Roth's argument and by improving the bounds on 
Szemer\'edi's theorem, Gowers~\cite{Gowers} defined the $U^s$-uniformity norms. Gowers' original 
definition works in the setting of the 
integers, and below 
we state the finite field version.
For $s\geq 1$,
the $s$-uniformity norm of a 
function $f\colon \mathbb{F}_p^n\to\mathbb{C}$ is defined as
\[
\norm{f}_{U^s}
=\left(\E_{x,h_1,\ldots,h_s\in\mathbb{F}_p^n}\left[\prod\limits_{w\in\{0,1\}^s}C^{\card{w}}f(x+\sum\limits_{i=1}^{s}w_ih_i)\right]\right)^{1/2^s},
\]
where $\card{w} = w_1+\ldots+w_s$ and $C^{j}$ is the conjugate operator for odd $j$ and the identity for even $j$. Using these 
norms, Gowers showed that if $A\subseteq \mathbb{F}_p^n$ has noticeable density and the normalized indicator $f_A = 1_A - \mu(A)$ has small $(k-1)$-uniformity norm, then $A$ 
contains many $k$-term arithmetic progressions (analogously to the case that $k=3$ and $f_A$ has all small Fourier coefficients). 
To proceed by a density increment strategy, Gowers then studies the case that $f_A$ has a noticeable uniformity norm, which is significantly more challenging. Using a combination of combinatorial techniques and powerful tools from additive combinatorics Gowers shows a local inverse theorem: there exists an affine subspace $W\subseteq \mathbb{F}_p^n$ of codimension $O_{k,p}(1)$ on which $f_A$ is correlated with a function of form $e^{2\pi{\bf i} \phi(x)}$, where $\phi\colon\mathbb{F}_p^n\to\mathbb{F}_p$ is a degree-$(k-2)$ polynomial. This allows him to conclude that there exists a subspace $W'\subseteq \mathbb{F}_p^n$ of dimension $n^{\Omega_{k,p}(1)}$ in which $A$ is noticeably denser. 

Since the work of Gowers, the uniformity norms have become an indispensable tool in additive combinatorics and number theory, and their study has received much attention, see for example~\cite{BergelsonTaoZiegler2010,TaoZiegler2012LowChar,GreenTaoZiegler2012UsPlus1,Manners2018Quantitative,Peluse2019Polynomial,Milicevic2023Inverse,LengSahSawhney2024Inverse,LengSahSawhney2024Szemeredi}. 
\subsection{Constraint Satisfaction Problems.} 
Our next example comes from the field of theoretical computer science, and more specifically from the area of hardness of approximation. 
\subsubsection{Decision and Optimization Problems.}
\begin{definition}
    Let $\Sigma$ be a finite set, and let $k\in\mathbb{N}$. A $k$-ary predicate over $\Sigma$ is a map $P\colon \Sigma^k\to\{0,1\}$.
\end{definition}
We often refer to the set $\Sigma$ as the alphabet
of the predicate, and to the parameter $k$ as its
arity. With any collection $\mathcal{P}$ of $k$-ary predicates over $\Sigma$ we associate the computational problem $\mathcal{P}$-CSP, defined as follows.
\begin{definition}
    For a collection of $\mathcal{P}$ of $k$-ary predicates over $\Sigma$, an 
    instance $\inst$ of $\mathcal{P}$-CSP consists of 
    a set of variables $X = \{x_1,\ldots,x_n\}$ and a 
    set of constraints $E = \{e_1,\ldots,e_m\}$. Each 
    constraint $e_i$
    consists of $k$-variables 
    $x_{i_1},\ldots,x_{i_k}\in X$
    and a predicate $P_i\in \mathcal{P}$, and it is thought of as imposing the condition $P_i(x_{i_1},\ldots,x_{i_k}) = 1$.
\end{definition}
Let $\mathcal{P}$ be a collection of predicates, and let $\inst$ be an instance of $\mathcal{P}$-CSP. 
An assignment to $\inst$ is a labeling $A\colon X\to \Sigma$. 
We say an assignment $A$ satisfies a constraint $e_i$ given as 
$P_i(x_{i_1},\ldots,x_{i_k}) = 1$, if
$P_i(A(x_{i_1}),\ldots,A(x_{i_k})) = 1$. We define the value of the assignment $A$ as $\val_{\inst}(A) = \frac{\card{\{e_i\in E~|~A\text{ satisfies }e_i\}}}{\card{E}}$, and define the value of an instance $\inst$ as
$\val(\inst) = \max_{A\colon V\to\Sigma}\val_{\inst}(A)$. We consider the following computational problems:
\begin{enumerate}
    \item Decision-Satisfiability: given an instance $\inst$ of $\mathcal{P}$-CSP, determine if $\val(\inst) = 1$ or not. In words, design an algorithm that given an instance $\inst$ of $\mathcal{P}$-CSP, accepts if $\val(\inst)=1$, and rejects otherwise.
    \item Gap-Maximization: for numbers $0\leq s\leq c\leq 1$, design an algorithm that given an instance $\inst$ of $\mathcal{P}$-CSP promised to either have $\val(\inst)\geq c$ or $\val(\inst)< s$, accepts in the former case and rejects in the 
    latter case. We often denote this promise problem by gap-$\mathcal{P}$-CSP$[c,s]$.
\end{enumerate}

Problems as in the first item above, often referred to as decision problems in the literature, are the basis of the theory of NP-completeness~\cite{Cook,Levin,Karp}. While the classical theory of NP-completeness goes far beyond the scope of constraint satisfaction problems, much effort has gone into characterizing collections $\mathcal{P}$ when the above task is computationally feasible. The dichotomy theorem of Zhuk and Bulatov~\cite{Zhuk,Bulatov} (which for a long time 
was a prominent outstanding conjecture) states that for any collection of predicates $\mathcal{P}$, the decision problem associated with $\mathcal{P}$-CSP can either be solved in polynomial time, or else is NP-hard.

Problems as in the second item above, often referred to as gap problems or optimization problems, are the basis of the theory of NP-hardness of approximation problems. 
This is precisely the type of problems addressed by the PCP theorem~\cite{FGLSS,AS,ALMSS}, which in these terms asserts that for $\Sigma = \{0,1\}$ and $k=3$, there is $s<1$ and a collection of $3$-ary predicates $\mathcal{P}$ such that gap-$\mathcal{P}$-CSP$[1,s]$ is NP-hard. While the theory of approximation is by now well-developed, we are still far from a complete understanding of the complexity of approximation of CSPs.

\subsubsection{The Almost Satisfiable Regime.}
The problem 
gap-$\mathcal{P}$-CSP$[c,s]$ turns out to exhibit a very different behavior depending on whether $c=1$  (the ``satisfiable regime''), or $c<1$ (typically $c=1-\eps$ for a small $\eps>0$, the ``almost satisfiable regime''). To demonstrate this we consider the $3$-Lin problem.

An instance of the $3$-Lin problem $\inst$ consists of a collection of variables $X = \{x_1,\ldots,x_n\}$ that are supposed to be assigned values from $\mathbb{F}_p$, and a collection of equations of the form $ax_i+bx_j+cx_k = d$ where $x_i,x_j,x_k\in X$ and $a,b,c,d\in\mathbb{F}_p$ are  constants (it is easy to formalize this problem as a CSP as defined above). 
The goal is to find an assignment $A\colon X\to\mathbb{F}_p$ satisfying as many of the equations as possible. 
For this problem, it is easy to see that gap-$3$-Lin$[1,s]$ can be solved in polynomial time for any $s\leq 1$. Indeed, the Gaussian elimination algorithm from linear algebra can be used to determine if the given system of linear equations has a solution, i.e.~if $\val(\inst)=1$, or not. The situation is completely different for $c<1$, and the Gaussian elimination algorithm no longer works. It can be shown that a randomly chosen assignment satisfies in expectation $1/p$ fraction of the equations, so it is always the case that $\val(\inst)\geq 1/p$. Thus, gap-$3$-Lin$[c,1/p]$ can be solved in polynomial time for all $c<1$. It turns out that this is essentially the best one can do:
\begin{theorem}[H\aa stad~\cite{Hastad}]\label{thm:hastad}
    For any prime $p$ and $\eps>0$, the problem
    gap-$3$-Lin$[1-\eps,1/p+\eps]$
    is NP-hard.
\end{theorem}
H\aa stad's proof of~\Cref{thm:hastad} proceeds by a reduction from the PCP theorem, and in his analysis he uses discrete Fourier analysis (in fact, he uses a similar Fourier-analytic argument to the one in~\Cref{sec:intro_AP}). Fourier-analytic tools have since become ubiquitous in the area.

Subsequent research led to an almost complete understanding of the ``almost satisfiable regime'', at least assuming a complexity theoretic assumption known as the Unique-Games Conjecture~\cite{Khot} (which will not be discussed here; see~\cite{KhotICM,TrevisanUGC,BarakSteurer2014SOSUGC,Khot2019TwoTwoSurvey,Minzer2022MonotonicityTwoTwo} for more information). This culminated in a result by Raghavendra~\cite{Raghavendra}, who proved the following dichotomy-type result:
\begin{theorem}[Raghavendra]\label{thm:raghavendra}
    For any finite alphabet $\Sigma$, $k\in\mathbb{N}$, a collection of $k$-ary predicates $\mathcal{P}$ over $\Sigma$, $c\in [0,1]$ and $\eps>0$, there exists $s\in [0,1]$ such that the following holds:
    \vspace{-1ex}
    \begin{enumerate}
        \item Algorithm: there is a polynomial time algorithm solving gap-$\mathcal{P}$-CSP$[c-\eps,s]$.
        \item Hardness: assuming the Unique-Games conjecture, for all $\delta>0$ 
        gap-$\mathcal{P}$-CSP$[c-\eps,s+\delta]$
        is NP-hard.
    \end{enumerate}
\end{theorem}

\subsubsection{Dictatorship Tests.}
The proof of~\Cref{thm:raghavendra} 
also proceeds by a reduction and uses discrete Fourier analysis. A key component in this result is a 
\emph{dictatorship test}. Roughly speaking, a $(c,s)$-dictatorship test for a predicate $P\colon \Sigma^k\to\{0,1\}$ is a distribution $\mu$ over $\Sigma^k$ such that the following holds for large enough $n$:
\vspace{-1ex}
\begin{enumerate}
    \item Completeness: if $f\colon \Sigma^n\to \Sigma$ is a dictatorship function, namely, it is a function of the form $f(x) = x_i$ for some $i\in\{1,\ldots,n\}$, then
    \[
    \E_{(x_1,\ldots,x_k)\sim \mu^{\otimes n}}[P(f(x_1),\ldots,f(x_k))]\geq c.
    \]
    Here, the distribution $\mu^{\otimes n}$ is the 
    distribution over $(\Sigma^n)^k$ where for each $i\in\{1,\ldots,n\}$, $(x_1(i),\ldots,x_k(i))$ is sampled independently according to $\mu$.
    \item 
    Soundness: if $f\colon \Sigma^n\to\Sigma$ is a function that is quasi-random with respect to dictatorships, then
    \[
    \E_{(x_1,\ldots,x_k)\sim \mu^{\otimes n}}[P(f(x_1),\ldots,f(x_k))]\leq s+o(1).
    \]
    By ``quasi-random with respect to dictatorships'' 
    we mean that for each $a\in\Sigma$, the function 
    $f_a\colon \Sigma^n\to\{0,1\}$
    defined as $f_a(x) = 1_{f(x) = a}$ has 
    that the individual influences $
    I_i[f_a; \mu_r^{\otimes n}] = \Pr_{x,x'\sim\mu_r^{\otimes n}}[f_a(x)\neq f_a(x')~|~x_j=x_j'~\forall j\neq i]$ are $o(1)$
     for all $i=1,\ldots,n$ and $r=1,\ldots,k$, where $\mu_r$ is the marginal distribution of $\mu$ on coordinate $r$.
    \end{enumerate}
In the proof of~\Cref{thm:raghavendra} Raghavendra gives an approximation algorithm based on semi-definite programming. He then considers integrality gaps for this algorithm -- namely instances which are ``hardest'' for the algorithm, and shows how to construct a dictatorship test $\mu$ with parameters nearly matching the performance of the algorithm.

A subtle point that initially seems very minor is that Raghavendra must ensure that the support of the distribution $\mu$, denoted by $\supp(\mu)$, is the entirety of $\Sigma^k$. Ultimately, this leads to the completeness parameter being $c-\eps$ (instead of $c$) in both of the items in~\Cref{thm:raghavendra}. This fact is necessary in the analysis of the dictatorship test $\mu$ that he constructs. Roughly speaking, after arithmetizing the predicate $P$, 
the analysis of the dictatorship tests boils down to understanding expectations of the form
\begin{equation}\label{eq:main_kwise_cor}
\E_{(x_1,\ldots,x_k)\sim \mu^{\otimes n}}[f_1(x_1)\cdots f_{k}(x_k)]
\end{equation}
for some $1$-bounded functions $f_1,\ldots,f_k$ arising from $f$ (in fact, each $f_i$ is $f_{a_i}$ for some $a_i\in \Sigma$). Raghavendra argues that 
almost all the contribution to this expectation comes from the low-degree Fourier part of the functions $f_i$, at which point he can replace the $f_i$ with their low-degree part. The argument is then finished by appealing to the invariance principle of Mossel, O'Donnell and Oleszkiewicz~\cite{MOO}, which relates the 
correlation of the low-degree parts to a similar expectation in Gaussian space.

To argue that essentially all contribution to
this expectation comes from the low-degree parts
of the functions Raghavendra uses a result of Mossel~\cite{Mossel}, which is the place where the fact that
$\supp(\mu) = \Sigma^k$ is used crucially.

\subsubsection{The Satisfiable Regime.}\label{sec:satisfiable_regime}
Results in the satisfiable regime
are much more rare in the literature, and we are far from having a complete understanding of the complexity of gap-$\mathcal{P}$-CSP$[1,s]$ for general $\mathcal{P}$. In fact, this is well understood only for very few predicates. An example is the $3$-SAT problem, which is defined by the $8$ predicates $\{P_{a,b,c}\colon \{0,1\}^{3}\to\{0,1\}~|~a,b,c\in\{0,1\}\}$ given as $P_{a,b,c}(x,y,z) = 1-1_{x=a,y=b,z=c}$. It is shown in~\cite{Hastad} that gap-$3$-SAT$[1,s]$ can be solved in polynomial time for $s = \frac{7}{8}$ and is NP-hard for any $s\geq \frac{7}{8}+\eps$.

The satisfiable regime remains rather poorly understood even if one is willing to assume conjectures in the spirit of the Unique-Games Conjecture, such as the Rich $2$-to-$1$ Games Conjecture~\cite{BKM}.
This is because essentially all hardness 
of approximation results are proved
by constructing a dictatorship test as above (as is the case in~\Cref{thm:raghavendra}), 
and to address the satisfiable regime one must come up with dictatorship tests with $c=1$.
In particular, the support of $\mu$ must be contained in $P^{-1}(1)$, and we can no longer afford to modify the distribution $\mu$ so that $\supp(\mu)=\Sigma^k$ entirely. We must work with $\mu$ as it is.

\section{The Main Analytical Questions.}
The former discussion naturally leads to the following question:
\begin{question}\label{question:basic_kwise_correlations}
    For what distributions $\mu$ over $\Sigma^k$ is the following assertion true: for all $1$-bounded functions $f_1,\ldots,f_k\colon \Sigma^n\to\mathbb{C}$ such 
    that at least one of the $f_i$ is essentially only on degrees higher than $d$, we have that
    \[
    \card{\E_{(x_1,\ldots,x_k)\sim\mu^{\otimes n}}[f_1(x_1)\cdots f_k(x_k)]}\rightarrow_{d\rightarrow \infty} 0.
    \]
\end{question}
Here and throughout, we will often use the somewhat 
informal term ``essentially on degrees higher than $d$''. By that, we mean that if we expand $f_i$ according to its Efron-Stein decomposition, only a very 
small amount of mass remains on degrees lower than $d$.
A typical example is a function of the form $f_i = (\mathrm{I}-\mathrm{T}_{1-\eps/d})g_i$, where $g_i$ is a $1$-bounded
function, $\mathrm{I}$ is the identity operator, and 
$\mathrm{T}_{1-\eps/d}$ is the standard noise operator,
defined as follows.
\begin{definition}
    For a distribution $\nu$ over $\Sigma$ and a point 
    $x\in \Sigma^n$, a sample from the distribution $\mathrm{T}_{1-\eps,\nu} x$ is drawn by taking for each $i\in\{1,\ldots,n\}$ independently $y_i = x_i$ with probability $1-\eps$, and else sampling $y_i\sim \nu$. When $\nu$ is clear from context we 
    often omit it from notation.
    Abusing notations, we also think of 
    $\mathrm{T}_{1-\eps}$ as an averaging operator, 
    where for $f\colon \Sigma^n\to\mathbb{C}$ we define
    $\mathrm{T}_{1-\eps}f(x) = \E_{y\sim \mathrm{T}_{1-\eps}x}[f(y)]$.
\end{definition}

As discussed earlier, Mossel~\cite{Mossel} proved that the assertion of~\Cref{question:basic_kwise_correlations} holds if $\supp(\mu) = \Sigma^k$. In fact, he showed that it is suffices that $\supp(\mu)$ is \emph{connected}. By that, we mean that the graph $G_{\mu}$ whose vertices are $\supp(\mu)$,
and $(a_1,\ldots,a_k),(b_1,\ldots,b_k)\in\supp(\mu)$
are adjacent if they differ in a single coordinate, 
is connected.

\subsection{The Case of No Abelian Embeddings.}
In an effort to make progress on the complexity of approximation of CSPs in the satisfiable regime, the 
authors of~\cite{CSPs1} hypothesize an answer to~\Cref{question:basic_kwise_correlations}. 
To state it we require the notion of \emph{Abelian embeddings}, which is central to the discussion.
\begin{definition}\label{def:Abelian_embed}
    Let $k\in\mathbb{N}$, let $\Sigma_1,\ldots,\Sigma_k$ be finite alphabets
    and let $\mu$ be a distribution over $\Sigma_1\times\ldots\times\Sigma_k$ 
    such that for each $i$, $\supp(\mu_i) = \Sigma_i$.
    We say that $\mu$ admits an Abelian embedding
    if there is an Abelian group $(A,+)$ and maps
    $\sigma_i\colon \Sigma_i\to A$ not all constant 
    such that    $\sigma_1(x_1)+\ldots+\sigma_k(x_k) = 0_A$ for all 
    $(x_1,\ldots,x_k)\in\supp(\mu)$.
\end{definition}
In words, a distribution is said to admit an Abelian embedding if the alphabet symbols can be (non-trivially) labeled by
elements of an Abelian group $(A,+)$, such that under this labeling, $\supp(\mu)$ is contained in the solution space of 
some linear equation over $A$. The hypothesis of~\cite{CSPs1} reads:

\begin{hypothesis}\label{hyp:no_Abelian}
    Suppose that $\mu$ is a distribution over $\Sigma^k$
    that does not admit any Abelian embedding. Then 
    the assertion of~\Cref{question:basic_kwise_correlations} 
    holds.
\end{hypothesis}

We remark that if $\mu$ admits an Abelian embedding, 
then the assertion of~\Cref{question:basic_kwise_correlations} fails. 
Indeed, take an Abelian group $(A,+)$ and 
maps $\sigma_1,\ldots,\sigma_k$ as in~\Cref{def:Abelian_embed}, and choose non-trivial characters
$\chi_1,\ldots\chi_n\in\hat{A}$. Define 
the functions $f_1,\ldots,f_k$ as 
$f_i(x_i)
=\prod\limits_{j=1}^{n}\chi_j(\sigma_i(x_i(j)))$.
Then for all $(x_1,\ldots,x_k)\in \supp(\mu^{\otimes n})$ we have
\[
\prod\limits_{i=1}^{k}f_i(x_i)
=\prod\limits_{i=1}^{k}
\prod\limits_{j=1}^{n}\chi_j(\sigma_i(x_i(j)))
=
\prod\limits_{j=1}^{n}
\prod\limits_{i=1}^{k}\chi_j(\sigma_i(x_i(j)))
=
\prod\limits_{j=1}^{n}\chi_j(\sum\limits_{i=1}^{k}\sigma_i(x_i(j)))
=
\prod\limits_{j=1}^{n}\chi_j(0_A)
=1,
\]
and in particular the expectation of $\prod\limits_{i=1}^{k}f_i(x_i)$ over $\mu^{\otimes n}$ 
has a large absolute value. Also, taking $i$ such that $\sigma_i$ is not constant, it can 
be checked that almost all of the
mass of $f_i$ is on degrees $\Theta(n)$.
This example means that, if true,~\Cref{hyp:no_Abelian} 
would give a complete answer to~\Cref{question:basic_kwise_correlations}.

\subsection{The Case of Abelian Embeddings.}
Recall that in Section~\ref{sec:AP} we discussed $k$-wise correlations of the form~\eqref{eq:main_kwise_cor} where the
distribution $\mu$ is uniform over $(x,x+a,x+2a,\ldots,x+(k-1)a)$ over all $x,a\in\mathbb{F}_p$. Observe that this $\mu$ does admit Abelian 
embeddings; for example, one can take 
$\sigma_i\colon \mathbb{F}_p\to \mathbb{F}_p$
defined as $\sigma_1(x) = x$, $\sigma_2(y) = -2y$, 
$\sigma_3(z) = z$ and $\sigma_i\equiv 0$ for $i>3$.
Thus,~\Cref{hyp:no_Abelian} does not apply for $\mu$. This raises the question: what can we say about the inverse problem for correlations such as~\eqref{eq:main_kwise_cor} when $\mu$ does have Abelian embeddings?

A quick inspection shows that for the problem to 
have any meaningful answer we must make some mild assumption on $\mu$. For instance, if $\mu$ is only supported on tuples of the form $(x,\ldots,x)\in \Sigma^k$, then essentially nothing of interest can be said about functions $f_1,\ldots,f_k$ that achieve high $k$-wise correlation with respect to $\mu$. A natural, fairly 
mild, property for $\mu$ turns out to be \emph{pairwise-connectedness}. 
Here and throughout, for $i,j\in \{1,\ldots,k\}$, $\mu_i$ 
and $\mu_{i,j}$ are the marginal distributions of $\mu$ on coordinate $i$, and on coordinates $i,j$, respectively.
\begin{definition}\label{def:pairwise_connected}
    Let $k\in\mathbb{N}$ and let 
    $\Sigma_{1},\ldots,\Sigma_k$ 
    be finite alphabets. 
    We say that a distribution 
    $\mu$ over $\Sigma_1\times\ldots\times\Sigma_k$
    is pairwise-connected if any distinct $i,j\in\{1,\ldots,k\}$, the bipartite graph $G_{i,j}$,
    whose sides are $\Sigma_i$, $\Sigma_j$ and whose edge set is $\supp(\mu_{i,j})$, is connected.
\end{definition}

It can be checked that for a distribution $\mu$, 
the following chain of implications holds:
\[
\mu\text{ is connected}~\Longrightarrow~
\mu\text{ has no Abelian embeddings}~\Longrightarrow~
\mu\text{ is pairwise-connected},
\]
and so among the $3$ notions, being pairwise-connected is the mildest property. It can be seen that all of the
distributions we considered so far (including those arising in the context of arithmetic progressions) 
are pairwise-connected.
The following question asks for a characterization of $1$-bounded functions $f_1,\ldots,f_k$
achieving noticeable correlation with respect to the $k$-wise correlation defined by $\mu$, in the case it is pairwise-connected:
\begin{question}\label{question:pairwise}
    Let $\mu$ be a pairwise-connected distribution over $\Sigma^k$. Suppose $f_1,\ldots,f_k\colon \Sigma^n\to\mathbb{C}$ are $1$-bounded functions 
    such that
    \[
    \card{\E_{(x_1,\ldots,x_k)\sim\mu^{\otimes n}}[f_1(x_1)\cdots f_k(x_k)]}\geq \eps,
    \]
    where $\eps>0$. What can we say about the structure of the functions $f_i$?
\end{question}
Below, after discussing some results, we state a natural candidate for the structure of functions as in~\Cref{question:pairwise}.

We remark that the setting of~\Cref{question:pairwise} generalizes the
setting of~\Cref{question:basic_kwise_correlations}, 
so the answer must include the possibility that $f_i$ are correlated with low-degree functions.
For $k=3$, the setting of~\Cref{question:pairwise} 
includes within it the setting of arithmetic progressions of length $3$, so the answer must include the possibility that $f_i$ are correlated with some Fourier character coming from an Abelian group (which is not clear a priori, since now $\Sigma$
is a completely abstract set, void of any algebraic structure). More generally, for large $k=2^s$ the setting of~\Cref{question:pairwise} includes within it $k$-term arithmetic progressions as well as $U^s$ uniformity norms. Thus, the answer must include the possibility that $f_i$ is correlated with
functions arising in the inverse theorem for Gowers uniformity norms as in~\cite{TaoZiegler}. As 
the inverse theorem for Gowers uniformity norms gets
significantly more challenging as $s$ increases, it
is natural to expect that~\Cref{question:pairwise}  gets more challenging as $k$ increases. 

\section{Inverse Theorems for $k$-Wise Correlations: Statements.}\label{sec:statements}
In this section we discuss results towards the 
resolution of~\Cref{question:basic_kwise_correlations} 
and~\Cref{question:pairwise}, and in~\Cref{sec:proofs} we give an informal
description of the proofs.
\subsection{The Case of $3$-ary Distributions with No Abelian Embeddings.}\label{sec:case_noab}
The paper~\cite{CSPs1} raised~\Cref{hyp:no_Abelian} 
and proved it is correct in the case $k=3$ and the 
distribution $\mu$ is a \emph{union of matchings}.
By that, we mean 
that for each $x$, there is a perfect matching $M_x\colon \Sigma\to \Sigma$ such that 
$\supp(\mu) = \bigcup_{x\in \Sigma}\{(x,y,M_x(y))~|~y\in\Sigma\}$. Their result reads 
as follows:
\begin{theorem}\label{thm:union of matchings}
    For all  $\alpha,\eps>0$ there exist $\delta>0$ and $d\in\mathbb{N}$
    such that the following holds. Suppose $\mu$ is a distribution over 
    $\Sigma^3$ satisfying:
    \vspace{-1ex}
    \begin{enumerate}
    \item The probability of each atom in $\mu$ is 
    at least $\alpha$.
    \item The distribution $\mu$ has no Abelian embeddings and is a union of matchings.
    \end{enumerate}
    Then if 
    $f,g,h\colon \Sigma^n\to\mathbb{C}$ are 
    $1$-bounded functions satisfying 
    $\card{\E_{(x,y,z)\sim \mu^{\otimes n}}[f(x)g(y)h(z)]}\geq \eps$, then there exists a function 
    $L\colon (\Sigma^n,\mu_1^{\otimes n})\to\mathbb{C}$
    of degree at most $d$ and $2$-norm equal to $1$ 
    such that $\card{\inner{f}{L}}\geq\delta$.
\end{theorem}

While there is nothing inherently interesting about the 
``union of matchings'' condition, it is a natural case to consider from a technical point of view. The proof of~\Cref{thm:union of matchings} involves a certain 
procedure called ``the path trick'' that, in a sense, enriches the distribution $\mu$. The case 
of union of matchings can be seen as sort of limiting point of that process.

\Cref{thm:union of matchings} has been subsequently improved in~\cite{CSPs2}, who removed 
the ``union of matchings'' condition. Their result reads as follows:
\begin{theorem}\label{thm:3CSP}
    For all  $\alpha,\eps>0$ there exist $\delta>0$ and $d\in\mathbb{N}$
    such that the following holds. Suppose $\mu$ is a distribution over 
    $\Sigma^3$ satisfying:
    \vspace{-1ex}
    \begin{enumerate}
    \item The probability of each atom in $\mu$ is 
    at least $\alpha$.
    \item The distribution $\mu$ has no Abelian embeddings.
    \end{enumerate}
    Then if 
    $f,g,h\colon \Sigma^n\to\mathbb{C}$ are 
    $1$-bounded functions satisfying 
    $\card{\E_{(x,y,z)\sim \mu^{\otimes n}}[f(x)g(y)h(z)]}\geq \eps$, then there exists a function 
    $L\colon (\Sigma^n,\mu_1^{\otimes n})\to\mathbb{C}$
    of degree at most $d$ and $2$-norm equal to $1$ 
    such that $\card{\inner{f}{L}}\geq\delta$.
\end{theorem}

The proof of~\Cref{thm:3CSP} uses some of the components of the proof of~\Cref{thm:union of matchings}, but does not reduce to the case therein. 
Instead, the proof proceeds by induction on $n$, the number of coordinates, via a sort of tensorization argument. 

\Cref{thm:3CSP} resolves~\Cref{hyp:no_Abelian} for $k=3$, and it is natural to next consider the case of $k=4$. This case turns out
to be significantly more difficult, and we do not know
how to reduce it to the $k=3$ case. 
A natural attempt at such a reduction proceeds as follows:
suppose we have $1$-bounded functions $f_1,\ldots,f_4\colon \Sigma^n\to\mathbb{C}$, 
and let $\mu$ be a distribution over $\Sigma^4$. 
Consider the distribution $\mu'$ over $(\Sigma')^3$
where $\Sigma' = \Sigma^2$, defined as follows:
\vspace{-1ex}
\begin{enumerate}
    \item Sample $(x,y,z,w)\sim\mu$.
    \item Sample $(x',y',z',w')\sim \mu$ conditioned on $w'=w$.
    \item Output $((x,x'),(y,y'),(z,z'))$.
\end{enumerate}
With this distribution in mind, one may use the $1$-boundedness of $f_4$ and Cauchy-Schwarz to get that
\begin{align}\label{eq:reduct_4_to_3}
\card{\E_{(x,y,z,w)\sim\mu^{\otimes n}}[f_1(x)f_2(y)f_3(z)f_4(w)]}^2
&=
\card{\E_{w\sim \mu_4^{\otimes n}}[f_4(w)\E_{(x,y,z,w')\sim\mu^{\otimes n}}[f_1(x)f_2(y)f_3(z)~|~w'=w]]}^2\notag\\
&\leq \E_{w\sim \mu_4^{\otimes n}}[\card{\E_{(x,y,z,w')\sim\mu^{\otimes n}}[f_1(x)f_2(y)f_3(z)~|~w'=w]}^2]\notag\\
&=\E_{((x,x'),(y,y'),(z,z'))\sim {\mu'}^{\otimes n}}[F_1(x,x')F_2(y,y')F_3(z,z')],
\end{align}
where $F_1(x,x') = f_1(x)\overline{f_1(x')}$, 
$F_2(y,y') = f_2(y)\overline{f_2(y')}$ 
and 
$F_3(z,z') = f_3(z)\overline{f_3(z')}$. It can be shown
that if $f_i$ is high-degree for some $i\in \{1,2,3\}$,
then the same holds for $F_i$. Thus, the above inequality would be a reduction from the $k=4$ to the 
$k=3$ case had it been the case that $\mu'$ has no Abelian embeddings whenever $\mu$ does not. Alas, this
turns out to be false.

Even though the above candidate reduction fails, it is 
not completely useless. 
We note that the support of $\mu'$ contains within it a copy of $\mu$, namely 
the ``equal pairs'' tuples $((x,x),(y,y),(z,z))$ for all $(x,y,z)\in \supp(\mu)$, and furthermore these inputs have a noticeable mass $\alpha'$ (which is at least $\alpha^2$) in $\mu'$. Thus, to generate a sample according to $\mu'$, one could write 
$\mu' = \alpha' \mu'' + (1-\alpha')\mu'''$
where $\mu''$ is the distribution $\mu$ conditioned on
being on equal pairs, and $\mu'''$ is the distribution 
$\mu$ conditioned on being on non-equal pairs, and 
then with probability $\alpha'$ sample according to $\mu''$, and with probability $1-\alpha'$ sample according to $\mu'''$.
To sample according to $\mu^{\otimes n}$, one could sample $I\subseteq_{\alpha'} [n]$,
by which we mean we include each $i\in [n]$ in $I$ with probability $\alpha'$ independently, then sample the coordinates of $I$
as $((x,x),(y,y),(z,z))\sim{\mu''}^{I}$
and the coordinates of $\bar{I}$ as $((x',\tilde{x}),(y',\tilde{y}),(z',\tilde{z}))\sim{\mu'''}^{\bar{I}}$.
With that in mind one can write the right hand side of~\eqref{eq:reduct_4_to_3} as
\[
\mathop{\E}_{\substack{I\subseteq_{\alpha'}[n]\\((x',\tilde{x}),(y',\tilde{y}),(z',\tilde{z}))\sim{\mu'''}^{\bar{I}}}}
\left[\mathop{\E}_{((x,x),(y,y),(z,z))\sim{\mu''}^{I}}[(F_1)_{\bar{I}\rightarrow (x',\tilde{x})}(x,x)(F_2)_{\bar{I}\rightarrow (y',\tilde{y})}(y,y)
(F_3)_{\bar{I}\rightarrow (z',\tilde{z})}(z,z)]\right],
\]
where $(F_1)_{\bar{I}\rightarrow (x',\tilde{x})}\colon {\Sigma'}^I\to\mathbb{C}$ is the function $F_1$ where the coordinates of $\bar{I}$ have been fixed to 
$(x',\tilde{x})$, and similarly for $F_2$ and $F_3$. The inner expectation can now be
thought of as an expectation over $\mu$, so one 
could deduce from the $k=3$ case that $(F_1)_{\bar{I}\rightarrow (x',\tilde{x})}$ is correlated with a low-degree function for many restrictions $\bar{I}, (x',\tilde{x})$. What could we
say about $F_1$ is that case?

It is useful to consider the case the distribution $\mu$ is supported on arithmetic progressions. In that case $F_1,F_2,F_3$ could be viewed as multiplicative derivatives of $f_1,f_2,f_3$, 
and the information we gathered indicates that they 
tend to be low-degree functions under axis-parallel restrictions. Morally, low-degree functions could be thought of as constant functions. Thus, as functions with constant multiplicative derivatives must be a Fourier characters, one expects something similar to hold for $F_1$.  

At the present setting though, this last assertion does not make complete sense. For once, we do not even have a group in the setting of a general $\mu$ as above. This naturally leads to~\Cref{question:pairwise}, with the hope that the answer therein will be useful in the above attempt for the $k=4$ case (which is indeed the case).

\subsection{The Case of $3$-ary Distributions with Abelian Embeddings.}
We now discuss results regarding~\Cref{question:pairwise} in the case 
that $k=3$. There is a significant 
difference depending on if the distribution $\mu$ admits an Abelian embedding into an infinite group
or not, and we use the following definition:
\begin{definition}
    We say a distribution $\mu$ over $\Sigma_1\times\ldots\Sigma_k$ admits $(\mathbb{Z},+)$-embeddings if there are maps
    $\sigma_1,\ldots,\sigma_k\colon \Sigma\to\mathbb{Z}$ 
    not all constant such that 
    $\sigma_1(x_1)+\ldots+\sigma_k(x_k) = 0$
    for all $(x_1,\ldots,x_k)\in\supp(\mu)$.
\end{definition}

We note that whenever a distribution $\mu$ admits a 
$(\mathbb{Z},+)$-embeddings, it also admits embeddings 
into arbitrarily large finite Abelian groups (taking $\sigma_i$ modulo $p$ for sufficiently large $p$, for example). The reverse is not true, though, and there are distributions that admit Abelian embeddings but have no $(\mathbb{Z},+)$-embeddings. For example,
the uniform distribution over $\{(x,x+a,x+2a)~|~x\in\mathbb{F}_p, a\in\{0,1,2\}\}$ admits Abelian embeddings but not $(\mathbb{Z},+)$-embeddings, whereas the uniform distribution over $\{(x,x+a,x+2a)~|~x\in\mathbb{F}_p, a\in\{0,1\}\}$ 
admits $(\mathbb{Z},+)$-embeddings.

To appreciate the difference between the case that $\mu$ has $(\mathbb{Z},+)$-embeddings and the case it does not, it is useful to inspect the recipe given below~\Cref{hyp:no_Abelian} for functions achieving noticeable $3$-correlations. Therein one 
could take $(A,+)$ to be any group which
$\mu$ embeds into. If $\mu$ has no $(\mathbb{Z},+)$-embeddings then there are only finitely many such $A$'s, 
and the functions $f_1,\ldots,f_k$ are discrete valued. On the other hand, if $(A,+) = (\mathbb{Z},+)$ then one can pick $\chi_j$ to
be any function of the form $\chi_j(a) = e^{2\pi{\bf i}\theta_ja}$ for $\theta_j\in (0,1)$. In particular, the choice of $\theta_j$ could depend on 
the number of variables $n$ (for example, it could be $\frac{1}{\sqrt{n}}$). This is a significantly richer 
class of examples, and they cannot be attributed to 
a single group of a fixed, constant size independent of $n$.

\subsubsection{The Subcase of No $(\mathbb{Z},+)$-embeddings.}
The paper~\cite{CSPs4} considered~\Cref{question:pairwise} in the case that 
$k=3$ and the distribution $\mu$ has no $(\mathbb{Z},+)$-embeddings. Their main technical result is a local inverse theorem:
\begin{theorem}\label{thm:csps4}
    For all $m\in\mathbb{N}$ there is a group $G$ of 
    size $O_{m}(1)$ such that for all $\eps,\alpha>0$ there exists $\delta>0$, such that if $\mu$ is a distribution 
    over $\Sigma^3$ in which $\card{\Sigma} = m$, the probability of each 
    atom is at least $\alpha$ and $\mu$ has no $(\mathbb{Z},+)$-embeddings, then there is a map $\sigma\colon \Sigma\to G$ such that the following holds. If $f,g,h\colon\Sigma^n\to\mathbb{C}$ are $1$-bounded functions such that $\card{\E_{(x,y,z)\sim \mu^{\otimes n}}[f(x)g(y)h(z)]}\geq \eps$, 
    then
    \[
    \Pr_{\substack{I\subseteq_{\delta}[n]\\ \tilde{x}\sim \mu_1^{\bar{I}}}}
    \left[\exists \chi\in \widehat{G}^{I}\text{ such that }\card{\inner{f_{\bar{I}\rightarrow\tilde{x}}}{\chi\circ\sigma^{\otimes I}}}\geq \delta\right]
    \geq \delta.
    \]
    Here, $f_{\bar{I}\rightarrow\tilde{x}}\colon \Sigma^I\to\mathbb{C}$ is the function $f$ when
    we restrict the coordinates of $\bar{I}$
    according to $\tilde{x}$, and 
    $\sigma^{\otimes I}\colon \Sigma^I\to G^I$
    is defined as 
    $\sigma^{\otimes I}(x) = (\sigma(x_i))_{i\in I}$.
\end{theorem}
In words,~\Cref{thm:csps4} asserts that if $f,g,h$
achieve noticeable $3$-wise correlation according to 
a distribution $\mu$ with no $(\mathbb{Z},+)$-embeddings, then after randomly restricting all but $\delta$ fraction of the coordinates, it is correlated 
with a character arising from some finite group $G$ of a fixed size.

To get a global inverse theorem they establish a
``restriction inverse theorems''~\cite{CSPs3, CSPs4}, giving structural information about a function $f$ from structural information about it under random restrictions:
\begin{theorem}\label{thm:csps4_restriction_inverse}
    Suppose $G$ is a finite Abelian group, $\sigma\colon \Sigma\to G$ is a map and $\nu$ is a distribution over $\Sigma$ in which the probability of each atom is at least $\alpha$.
    Then for all $\eps>0$ there are $d\in\mathbb{N}$ and $\delta>0$ such that if $f\colon\Sigma^n\to\mathbb{C}$ is a $1$-bounded function satisfying
    \[
    \Pr_{\substack{I\subseteq_{\eps}[n]\\ \tilde{x}\sim \nu^{\bar{I}}}}
    \left[\exists \chi\in \widehat{G}^{I}\text{ such that }\card{\inner{f_{\bar{I}\rightarrow\tilde{x}}}{\chi\circ\sigma^{\otimes I}}}\geq \eps\right]\geq \eps,
    \]
    then there exist $\chi\in\widehat{G}^n$ 
    and $L\colon(\Sigma^n,\nu^{\otimes n})\to\mathbb{C}$ with $\norm{L}_2=1$ and 
    ${\sf deg}(L)\leq d$ such that 
    $\card{\inner{f}{L\cdot \chi\circ\sigma^{\otimes n}}}\geq \delta$.
\end{theorem}
Combining~\Cref{thm:csps4,thm:csps4_restriction_inverse} gives an answer to~\Cref{question:pairwise} in the case $k=3$ and $\mu$ has no $(\mathbb{Z},+)$-embeddings:
\begin{theorem}\label{thm:csps4_global}
    In the setting of~\Cref{thm:csps4}, 
    for all $\eps>0$ there are $d\in\mathbb{N}$
    and $\delta>0$ such that if $f,g,h\colon \Sigma^n\to\mathbb{C}$ are $1$-bounded functions
    satisfying $\card{\E_{(x,y,z)\sim \mu^{\otimes n}}[f(x)g(y)h(z)]}\geq \eps$, then there exist $\chi\in\widehat{G}^n$ 
    and $L\colon(\Sigma^n,\mu_1^{\otimes n})\to\mathbb{C}$ with $\norm{L}_2=1$ and 
    ${\sf deg}(L)\leq d$ such that 
    $\card{\inner{f}{L\cdot \chi\circ\sigma^{\otimes n}}}\geq \delta$.
\end{theorem}

It is not clear though how to 
generalize the arguments of~\cite{CSPs4} to the 
case $\mu$ has $(\mathbb{Z},+)$-embeddings. The main issue is that the arguments in~\cite{CSPs4}
heavily rely on the fact that group $G$ can be 
identified by only looking at the distribution $\mu$. In particular, it cannot depend on the number of coordinates $n$. 

\subsubsection{The Pairwise-connected Case.}
The statements of~\cref{thm:csps4,thm:csps4_global,thm:csps4_restriction_inverse} 
still make some sense for $3$-ary, pairwise-connected distributions $\mu$ (namely, besides the fact that the size of $G$ depends only on $m$). In particular, one
is naturally led to the following candidate statement: in the setting of~\Cref{thm:csps4}, if $\mu$ is only assumed to be pairwise-connected (as opposed to having no $(\mathbb{Z},+)$-embeddings), then after random restriction it is correlated with some character function coming from $(\mathbb{Z},+)$.

We now make a few remarks. First, this candidate statement is false if we do not allow for random restrictions. In fact, we are not 
aware of any similar plausible statement that  
does not involve random restrictions, and this complicates matters. Second, once 
we allow $(\mathbb{Z},+)$-characters, the group structure becomes more of a red-herring, and it is better to
abstract out this structure in the language of \emph{product functions}.
\begin{definition}\label{def:prod_fn}
    We say a function $P\colon \Sigma^n\to\mathbb{C}$
    is a product function if there are univariate functions $P_1,\ldots,P_n\colon \Sigma\to\mathbb{C}$
    such that $P(x_1,\ldots,x_n) = P_1(x_1)\cdots P_n(x_n)$ for all $x_1,\ldots,x_n\in\Sigma$.
\end{definition}

With this definition, the main result of~\cite{CSPs6} is the following local inverse theorem:
\begin{theorem}\label{thm:csps6}
    For all $\eps,\alpha>0$ there exists $\delta>0$, such that the following holds for any pairwise-connected distribution $\mu$ 
    over $\Sigma^3$ in which the probability of each 
    atom is at least $\alpha$. If $f,g,h\colon\Sigma^n\to\mathbb{C}$ are $1$-bounded functions such that $\card{\E_{(x,y,z)\sim \mu^{\otimes n}}[f(x)g(y)h(z)]}\geq \eps$, 
    then
    \[
    \Pr_{\substack{I\subseteq_{\delta}[n]\\ \tilde{x}\sim \mu_1^{\bar{I}}}}
    \left[\exists P\colon\Sigma^I\to\mathbb{C}\text{ a product function with $\norm{P}_2\leq 1$ such that }\card{\inner{f_{\bar{I}\rightarrow\tilde{x}}}{P}}\geq \delta\right]\geq \delta.
    \]
\end{theorem}

The paper~\cite{CSPs6} also proves a version of~\Cref{thm:csps4_restriction_inverse} for product functions:
\begin{theorem}\label{thm:csps6_restriction_inverse}
    Suppose $\nu$ is a distribution over $\Sigma$ in which the probability of each atom is at least $\alpha$.
    Then for all $\eps>0$ there are $d\in\mathbb{N}$ and $\delta>0$ such that if $f\colon\Sigma^n\to\mathbb{C}$ is a $1$-bounded function satisfying
    \[
    \Pr_{\substack{I\subseteq_{\eps}[n]\\ \tilde{x}\sim \nu^{\bar{I}}}}
    \left[\exists P\colon\Sigma^I\to\mathbb{C}\text{ a product function with $\norm{P}_2\leq 1$ such that }\card{\inner{f_{\bar{I}\rightarrow\tilde{x}}}{P}}\geq \eps\right]\geq \eps,
    \]
    then there exist a $1$-bounded product function $P\colon \Sigma^n\to\mathbb{C}$ and $L\colon(\Sigma^n,\nu^{\otimes n})\to\mathbb{C}$ with $\norm{L}_2=1$ and 
    ${\sf deg}(L)\leq d$ such that 
    $\card{\inner{f}{L\cdot P}}\geq \delta$.
\end{theorem}

Combining~\Cref{thm:csps6,thm:csps6_restriction_inverse}
answers~\Cref{question:pairwise} in the case $k=3$:
\begin{theorem}\label{thm:csps6_global}
    For all $\eps,\alpha>0$ there exist $d\in\mathbb{N}$ and $\delta>0$, such that the following holds for a pairwise-connected distribution $\mu$ 
    over $\Sigma^3$ in which the probability of each 
    atom is at least $\alpha$. If $f,g,h\colon\Sigma^n\to\mathbb{C}$ are $1$-bounded functions such that $\card{\E_{(x,y,z)\sim \mu^{\otimes n}}[f(x)g(y)h(z)]}\geq \eps$, 
    then there exist a $1$-bounded product function $P\colon \Sigma^n\to\mathbb{C}$ and 
    a function 
    $L\colon \Sigma^n\to\mathbb{C}$ with 
    $\norm{L}_2 = 1$ and ${\sf deg}(L)\leq d$
    such that $\card{\inner{f}{L\cdot P}}\geq \delta$.
\end{theorem}

\subsection{The Case of $k$-ary Distributions with No Abelian Embeddings.} 
\Cref{thm:csps6_global} turns out to be sufficiently
strong to complete the argument outlined in~\Cref{sec:case_noab}
for the case that $k=4$ and $\mu$ has no Abelian embeddings~\cite{CSPs7}. In fact the paper~\cite{CSPs7} shows that it is sufficiently 
strong to address all constant arities $k$:
\begin{theorem}\label{thm:csps7}
    \Cref{hyp:no_Abelian} is true, and so the answer to~\Cref{question:basic_kwise_correlations} is 
    positive if and only if $\mu$ admits no Abelian embeddings.
\end{theorem}

\subsection{The Case of $k$-ary Distributions with Abelian Embeddings?} 
At present time,~\Cref{question:pairwise} is open
for all $k\geq 4$, and we believe it is a very challenging and interesting problem.~\Cref{thm:csps6_global} suggests a 
natural and plausible answer for all $k$, and to state it we define 
the notion of combinatorial low-degree functions:
\begin{definition}
    We say a function $P\colon \Sigma^n\to\mathbb{C}$
    is a combinatorial degree-$k$ function if 
    for each $T\in\binom{[n]}{k}$ there is a function $P_T\colon \Sigma^T\to\mathbb{C}$ such that 
    $P(x) = \prod\limits_{T\in\binom{[n]}{k}}P_T(x_T)$ for all $x\in\Sigma^n$, where $x_T\in \Sigma^T$ is the vector whose entries are the coordinates of $x$ from $T$.
\end{definition}

\begin{conjecture}\label{conj:general_inverse}
    For all $k\in\mathbb{N}$ there exists $k'\in\mathbb{N}$, such that for all $\eps,\alpha>0$
    there are $d\in\mathbb{N}$ and $\delta>0$ such that the following holds.
    Suppose $\mu$ is a pairwise-connected distribution 
    over $\Sigma^k$ in which the mass of each atom is at least $\alpha$. If 
    $f_1,\ldots,f_k\colon \Sigma^n$ are $1$-bounded
    functions such that 
    $\card{\E_{(x_1,\ldots,x_k)\sim\mu^{\otimes n}}[f_1(x_1)\cdots f_k(x_k)]}\geq \eps$, 
    then there is a combinatorial degree-$k'$ function $P\colon \Sigma^n\to\mathbb{C}$ with 
    $2$-norm equal to $1$, and $L\colon\Sigma^n\to\mathbb{C}$
    with $2$-norm equal to $1$ and ${\sf deg}(L)\leq d$ such that 
    $\card{\inner{f_1}{L\cdot P}}\geq \delta$.
\end{conjecture}
It would be very interesting to resolve~\Cref{conj:general_inverse} either in 
the positive or in the negative. If~\Cref{conj:general_inverse} is true, we believe that its proof will require developing
combinatorial analogs of tools from additive
combinatorics, which will be of independent interest. 
Subsequently, it will be interesting to develop density
increment arguments that work with combinatorial 
low-degree functions, and use it to deduce results 
that are beyond the scope of the uniformity norms 
(we give a few examples in~\Cref{sec:applications}). 
If~\Cref{conj:general_inverse} is false it would be 
interesting to examine counter-examples and modify~\Cref{conj:general_inverse} to a plausible statement that is still useful.

\section{Inverse Theorems for $k$-Wise Correlations: Proofs.}\label{sec:proofs}
In this section we outline some of the ideas in 
the proofs of the statements presented in~\Cref{sec:statements}.
\subsection{The Case of $3$-ary Distributions with No Abelian Embeddings.}

\subsubsection{The Reduction to Non-Abelian Groups: Proof of~\Cref{thm:union of matchings}}
The proof of~\Cref{thm:union of matchings} proceeds via 
a reduction to a problem in non-Abelian Fourier analysis. To explain the argument it will be convenient to have a different notation for the alphabet of each input, and we denote the alphabets of the first, second and third inputs by $\Sigma$, $\Gamma$ and $\Phi$ respectively. 

We now explain how to (separately) associate each element in $\Sigma,\Gamma,\Phi$ with an element in $S_{\Sigma}$, the symmetric group over $\Sigma$. For $\Sigma$, the association is given 
by $x\rightarrow M_x$ when we view the matching $M_x$ as a permutation. For $\Gamma$ and $\Phi$ this association is done via propagation. 
Take $y^{\star}\in \Gamma$ arbitrarily and identify it an arbitrary chosen permutation $\pi_{y^{\star}}\in S_{\Sigma}$. Next, inspect the graph $G_{2,3}$ associated with
$\mu$, and propagate the association via adjacency.
Namely, for each $x$ we associate the neighbor $z$ of $y^{\star}$ given as $z = M_x(y^{\star})\in \Phi$ with $\pi_{z} = M_x\circ \pi_{y^{\star}}$. Next, for each $z$ already associated with a permutation, 
we consider the neighbor $y = M_{x}^{-1}(z)\in \Gamma$ for each $x\in\Sigma$ and associate it with the permutation $\pi_y = M_x^{-1}\circ \pi_z$. We repeat this process for sufficiently many steps.\footnote{During this process it may be the case that a symbol is associated with two distinct permutations. Such cases are handled by operations called ``merges'', and ultimately boil down to identifying two symbols of $\Sigma$, $\Gamma$ or $\Phi$.}

It can be argued that since $G_{2,3}$ is connected, we will end up associating all
alphabet symbols with permutations, so our distribution 
$\mu$ naturally corresponds to some $3$-ary distribution over the permutation group $S_{\Sigma}$.
With some work
one can relate the expectation~\eqref{eq:main_kwise_cor} 
with a similar expectation over $S_{\Sigma}^n$, which can 
then be analyzed using elementary non-Abelian Fourier analysis.

\subsubsection{The Inductive Approach: Proof of~\Cref{thm:3CSP}.}
The high level approach in the proof of~\Cref{thm:3CSP} is inspired by classical tensorization arguments as in~\cite{Mossel}. 
To start, note that the assumption 
that $\mu$ has no Abelian embeddings can be equivalently stated as asserting that there is a constant $\lambda = \lambda(\mu)>0$ such that 
for all $1$-bounded functions 
$u\colon\Sigma\to\mathbb{C}$, 
$v\colon\Gamma\to\mathbb{C}$ 
and $w\colon \Phi\to\mathbb{C}$ with expectation $0$,
\begin{equation}\label{eq:base_case_CSPs2}
\card{\E_{(x,y,z)\sim \mu}[u(x)v(y)w(z)]}\leq 1-\lambda.
\end{equation}
Indeed, the upper bound of $1$ is trivial by the triangle inequality, and it is tight if and only if 
for all $(x,y,z)\in\supp(\mu)$ we have that
$u(x)v(y)w(z) = \theta$, where $\theta\in \mathbb{C}$ is some constant. Such an identity can be viewed as an Abelian embedding of $\mu$ to $([0,2\pi),+\pmod{2\pi})$, hence it does not exist if
$\mu$ has no Abelian embeddings.

With this in mind, it is natural to expect that 
if $f,g,h$ are functions of degree at least large $d$ as in~\Cref{hyp:no_Abelian}, then the univariate 
inequality above would tensorize and give us a bound
of $(1-\lambda)^d\rightarrow_{d\rightarrow \infty} 0$.
There are several issues with this approach, though:
\vspace{-1ex}
\begin{enumerate}
    \item A base case inequality such as~\eqref{eq:base_case_CSPs2} does not tensorize well. Specifically, the $1$-boundedness assumption is not well-preserved under 
    spectral/Fourier-analytic techniques. Instead, one needs a base case for functions bounded in $L_2$-norm.
    \item Tensorizing inequalities such as~\eqref{eq:base_case_CSPs2}, namely inequalities involving a product of $3$ functions, is far from being automatic. 
    This is in contrast to similar inequalities involving only two functions $u,v$, which can be stated in terms of eigenvalues of some matrix and thus tensorizes well (as was done by Mossel~\cite{Mossel}).
\end{enumerate}

To overcome these issues, the work~\cite{CSPs2} first establishes an alternative base case for univariate 
functions bounded in 
$L_2$-norm. Due to its technical nature we refer the reader to~\cite{CSPs2} for a precise statement of this base case, but remark that it is not simply the assertion that~\eqref{eq:base_case_CSPs2} holds 
for functions with $L_2$-norm at most $1$ (we do not know how to deduce such a base case from the assumption that $\mu$ has no Abelian embeddings). The base case of~\cite{CSPs2} is in fact a statement about a distribution $\mu'$ different from $\mu$, for which expectations of the form~\eqref{eq:main_kwise_cor} over $\mu$ can be upper bounded by similar expectations over $\mu'$. 

The main argument of~\cite{CSPs2} 
proceeds by induction on $n$ using the singular-value decomposition (SVD in short), asserting that $f\colon\Sigma^n\to\mathbb{C}$, 
$g\colon\Gamma^n\to\mathbb{C}$ 
and $h\colon\Phi^n\to\mathbb{C}$ with $2$-norm equal to $1$ can be written as
\[
f(x)
=\sum\limits_{i}\lambda_i u_i(x_1)f_i(x_2,\ldots,x_n),
\qquad
g(y)
=\sum\limits_{i}\gamma_i v_i(y_1)g_i(y_2,\ldots,y_n),
\qquad
h(z)
=\sum\limits_{i}\kappa_i w_i(z_1)h_i(z_2,\ldots,z_n),
\]
where each one of $\{u_i\}, \{f_i\}, \{v_i\}, \{g_i\}, \{w_i\}, \{h_i\}_i$ is an orthonormal system and $\lambda_i,\gamma_i,\kappa_i$ are non-negative real numbers satisfying $\sum\limits_i \lambda_i^2=\sum\limits_i \gamma_i^2=\sum\limits_i \kappa_i^2=1$. If each one of the SVDs involves only a single summand, then 
\[
\E_{(x,y,z)\sim\mu^{\otimes n}}[f(x)g(y)h(z)]
=
\E_{(x_1,y_1,z_1)\sim\mu}[u_1(x)v_1(y)w_1(z)]
\E_{(x,y,z)\sim\mu^{\otimes(n-1)}}[f_1(x)g_1(y)h_1(z)].
\]
To complete the inductive step the first expectation can be bounded using the base case, and the second expectation can be bounded using the inductive hypothesis. 

If the SVDs 
of $f,g$ and $h$ involve more than a 
single term, then there are many additional terms that need to be upper bounded, and the argument proceeds by considering two subcases. If there are SVDs so that essentially all of the mass
of $f$, $g$ and $h$ lies on a single component, then
one shows that the other terms do not interfere much, and one essentially only has to account for the contribution of the main term as above. Otherwise, it can be shown that the degrees of $f,g$ and $h$ all must be $\Theta(n)$, 
and one proceeds via a different inductive argument. One carefully chooses $F,G,H$ which are random linear combinations of $f_i$, $g_i$, $h_i$ respectively (these are $(n-1)$-variate functions), and shows
that the $3$-wise correlation of $f,g,h$ according to $\mu^{\otimes n}$ is at most $(1-\Omega(1))$ times the $3$-wise correlation of $F,G,H$ according to $\mu^{\otimes(n-1)}$. As $f,g,h$
are $\Theta(n)$ degree $F,G,H$ are also $\Theta(n)$ 
degree, so one can induct $\Omega(n)$ times, completing the proof.

\subsection{The Case of No $(\mathbb{Z},+)$-embeddings: Proof of~\Cref{thm:csps4}.}  By considering the set of equations defining Abelian embeddings as a linear system of equations, one can show that if $\mu$ admits no $(\mathbb{Z},+)$-embeddings, then all of its embeddings are equivalent to an embedding into $G=\prod\limits_{p\leq p_{{\sf max}, \ell\leq \ell_{{\sf max}}}}\mathbb{Z}_{p^{\ell}}$, 
where $p_{\sf max},\ell_{\sf max}$ depend only on $\card{\Sigma}$. This will be the group $G$ in the 
theorem, and one fixes a canonical embedding $(\sigma,\gamma,\phi)$ of $\mu$ into $G$. 
By that, we mean an embedding that encapsulates within it 
all of the Abelian embeddings of $\mu$.

The proof of~\Cref{thm:csps4} follows the spirit of~\Cref{thm:3CSP}, but there are many 
differences:
\vspace{-1ex}
\begin{enumerate}
    \item The base case inequality along the lines of~\eqref{eq:base_case_CSPs2} 
    cannot be true if $\mu$ has any Abelian embedding (indeed, applying a character on any Abelian embeddings gives a counter example). 
    Instead, one defines the space of univariate ``embedding functions'', which are functions 
    of the form $\chi\circ \sigma$, $\chi\circ\gamma$, $\chi\circ \phi$ for $\chi\in\hat{G}$, and requires a 
    base case inequality along the lines of~\eqref{eq:base_case_CSPs2} when at least one of the functions $u,v,w$ is far from being an embedding function. 
    \item In the inductive arguments, one considers a more refined degree notion for $f,g,h$, called the \emph{non-embedding degree}. Roughly speaking, one defines a basis for $L_2(\Sigma, \mu_1)$ (and similarly for $L_2(\Sigma, \mu_2)$ 
    and $L_2(\Sigma, \mu_3)$) consisting of (1) embedding functions, (2) functions orthogonal 
    to all embedding functions, and then tensorizes it to get a basis for $L_2(\Sigma^n,\mu_1^{\otimes n})$. One can then write $f_1$ (and similarly $f_2,f_3$) according to this basis, and consider
    separately the notions of (1) embedding degree (which is the number of functions from the first part of the basis), (2) the non-embedding degree (which is the number of functions from the second part of the basis).
\end{enumerate}
The proof of~\Cref{thm:csps4} proceeds via similar SVD-based inductive arguments using the notion of non-embedding degree instead of the standard notion of degree above. 
The conclusion from the inductive arguments is that the non-embedding degree of each one of $f,g,h$
must be small. Thus, after a random restriction their non-embedding degree becomes constant, meaning they are essentially functions over the group $G$! The $3$-wise correlation of the restrictions can now be studied using Fourier analysis over the group $G$, leading to the conclusion of~\Cref{thm:csps4}.

\subsection{The Pairwise-connected Case: Proof of~\Cref{thm:csps6}.}\label{sec:csps6} 
While the proof of~\Cref{thm:csps6} uses some ideas 
from the arguments discussed above, the overarching proof strategy is completely different. The main idea is to identify a norm, called \emph{the swap norm}, and show that:
\vspace{-1ex}
\begin{enumerate}
    \item 
it (morally) dominates all $3$-wise correlations with respect to a pairwise-connected distribution $\mu$, and 
\item it admits an inverse theorem in the spirit of~\Cref{thm:csps6}. 
\end{enumerate}
The definition of this norm is independent of $\mu$, meaning it simultaneously works for pairwise-connected $3$-ary distribution. This is reminiscent of the notion of ``Cauchy-Schwarz'' complexity~\cite{GowersWolf2011Linear}, asserting that the Gowers uniformity norms dominate correlations defined by collections of linear forms. 
\subsubsection{The Swap Norm.}
To define the swap norm we require the definition of the box-norm. Below, the notation $x\in_R\Sigma^n$ means that $x$ is sampled uniformly from $\Sigma^n$, and the notation $I\subseteq_{1/2}[n]$ means that each coordinate $i\in [n]$ is included in $I$ with probability $1/2$.
\begin{definition}
    Let $f_1,f_2,f_3,f_4\colon \Sigma^n\to\mathbb{C}$ be functions,
    and let $I\subseteq[n]$ be a set of coordinates. 
    The box form ${\sf box}_I(f_1,f_2,f_3,f_4)$ is defined as
    \[
    {\sf box}_I(f_1,f_2,f_3,f_4) = \mathop{\E}_{\substack{x,x'\in_R\Sigma^I\\ y,y'\in_R\Sigma^{\bar{I}}}}[f_1(x,y)f_2(x',y')\overline{f_3(x,y')}\overline{f_4(x',y)}].\]
    The box norm of a function $f\colon \Sigma^n\to\mathbb{C}$ with respect to $I\subseteq[n]$ is defined as 
    ${\sf box}_I(f) = {\sf box}(f,f,f,f)^{1/4}$.
\end{definition}

\begin{definition}
     Let $f_1,f_2,f_3,f_4\colon \Sigma^n\to\mathbb{C}$ be functions. 
    The swap form ${\sf swap}(f_1,f_2,f_3,f_4)$ is defined as
    \[
    {\sf swap}(f_1,f_2,f_3,f_4) = \E_{I\subseteq_{1/2}[n]}[{\sf box}_I(f_1,f_2,f_3,f_4)].
    \]
    The swap norm of a function $f\colon \Sigma^n\to\mathbb{C}$ is defined as ${\sf swap}(f) = {\sf swap}(f,\ldots,f)^{1/4}$.
\end{definition}

The box form is well known to have many useful properties such as the Cauchy-Schwarz-Gowers inequality~\cite{Gowers2007Hypergraph}. Being 
the average of box forms, the swap form also satisfies 
similar properties, such as:
\[
\card{{\sf swap}(f_1,f_2,f_3,f_4)}^2
\leq {\sf swap}(f_1,f_2,f_1,f_2)
{\sf swap}(f_3,f_4,f_3,f_4),
\qquad 
\card{{\sf swap}(f_1,f_2,f_1,f_2)}^2
\leq 
{\sf swap}(f_1)
{\sf swap}(f_2).
\]
These properties can be used to show that the swap norm is in fact a norm. The source of the name ``swap norm'' comes from the following identity. For $x,y\in\Sigma^n$, 
let $x\longleftrightarrow y$ be the distribution over $(u,v)\in\Sigma^n\times \Sigma^n$ where for each $i\in [n]$ independently, with probability $1/2$ we have $(u_i,v_i) = (x_i,y_i)$, and else we have $(u_i,v_i) = (y_i,x_i)$. With this notation, it can be easily verified by expanding the definition that 
\[
{\sf swap}(f_1,f_2,f_3,f_4)
=\mathop{\E}_{\substack{x,y\in_R\Sigma^n\\ (u,v)\sim (x\leftrightarrow y)}}\left[f_1(x)f_2(y)\overline{f_3(u)}\overline{f_4(v)}\right].
\]

The connection between $3$-wise correlations and the swap
norm comes from the following lemma, asserting that 
(under random restrictions) the swap norm of a function $f$ dominates $3$-wise correlations:
\begin{lemma}\label{lem:swap_dominates}
    Let $\mu$ be a pairwise-connected distribution 
    over $\Sigma^3$ in which the probability of each atom is at least $\alpha$, and write $\mu_1 = \alpha U + (1-\alpha)\nu$, where $\mu_1$ is the marginal distribution of $\mu$ on its first coordinate, $U$ is the uniform distribution over $\Sigma$ and 
    $\nu$ is some distribution over $\Sigma$. 
    Then there exists $C = C(\alpha)>0$ such that for all $1$-bounded functions 
    $f,g,h\colon \Sigma^n\to\mathbb{C}$ it holds that
    \[
    \card{\E_{(x,y,z)\sim \mu^{\otimes n}}[f(x)g(y)h(z)]}^{C(\alpha)}
    \leq
    \mathop{\E}_{\substack{I\subseteq_{\alpha}[n] \\ w\sim \nu^{\bar{I}}}}\left[
    {\sf swap}(f_{\bar{I}\rightarrow w})^4
    \right].
    \]
\end{lemma}

The proof of~\Cref{lem:swap_dominates} is by a symmetry argument involving several applications of the Cauchy-Schwarz inequality
(as well as random restrictions), and we do not give it 
here. 

\subsubsection{An Inverse Theorem for the Swap Norm.}
By an averaging argument, Lemma~\ref{lem:swap_dominates} asserts that if 
$\card{\E_{(x,y,z)\sim \mu^{\otimes n}}[f(x)g(y)h(z)]}\geq \eps$, then with noticeable probability over the choice of $I$ and $w$ we have that ${\sf swap}(f_{\bar{I}\rightarrow w})$ is 
noticeable. Thus,~\Cref{thm:csps6} boils down to the following \emph{inverse theorem
for the swap norm}:
\begin{theorem}\label{thm:swap_inverse_theorem}
    For all $\eps>0$ there exists $\delta>0$ such that the following holds. Suppose that 
    $f\colon \Sigma^n\to\mathbb{C}$ is a $1$-bounded 
    function such that ${\sf swap}(f)\geq \eps$. 
    Then
    \[
    \Pr_{\substack{I\subseteq_{\delta}[n]\\ \tilde{x}\in_R \Sigma^{\bar{I}}}}
    \left[\exists P\colon\Sigma^I\to\mathbb{C}\text{ a product function with $\norm{P}_2\leq 1$ such that }\card{\inner{f_{\bar{I}\rightarrow\tilde{x}}}{P}}\geq \delta\right]\geq \delta.
    \]
\end{theorem}

To get some intuition for~\Cref{thm:swap_inverse_theorem}, it is useful 
to think about the box norm and the following (rather elementary) inverse theorem for it.  If ${\sf box}_I(f)\geq \eps$, then by definition we may find $x'\in \Sigma^I$ and $y'\in \Sigma^{\bar{I}}$ such that
\[
\card{f(x',y')\mathop{\E}_{\substack{x\in_R\Sigma^I\\ y\in_R\Sigma^{\bar{I}}}}[f(x,y)\overline{f(x,y')}\overline{f(x',y)}]}
\geq {\sf box}_I(f)
\geq \eps.
\]
Using the $1$-boundedness of $f$ and defining the functions 
$g_I\colon \Sigma^I\to\mathbb{C}$
and
$h_{\bar{I}}\colon \Sigma^{\bar{I}}\to\mathbb{C}$
by $g(x) = f(x,y')$, $h(y) = f(x',y)$, 
we conclude that 
$\card{\inner{f}{g_Ih_{\bar{I}}}}\geq \eps$. 
In words, if $f$ has a large box norm with respect to $I\subseteq[n]$, then it is correlated 
with a product of $2$ functions, one depending only
on the coordinates of $I$ and the other depending only
on the coordinates of $\bar{I}$.

Returning to the swap norm, an averaging argument gives that if ${\sf swap}(f)\geq \eps$, then for $I\subseteq_{1/2}[n]$ with noticeable probability 
it holds that ${\sf box}_I(f)\geq \eps/2$. Applying the inverse theorem for the box norm, we get that 
for $I\subseteq_{1/2}[n]$, with noticeable probability
the function $f$ is correlated with a function of the form $g_Ih_{\bar{I}}$. It is easy to see that product functions as in~\Cref{def:prod_fn} are natural examples of 
functions $f$ satisfying this property. 
A closer inspection shows that functions of essentially low-degree may also satisfy this property (for example, the majority function). 
This means that 
deducing the product-function structure cannot
be straightforward and still requires some effort. 
In particular, the random-restriction part in~\Cref{thm:swap_inverse_theorem} is still essential.

A natural attempt at proving~\Cref{thm:swap_inverse_theorem} 
proceeds via an energy-increment argument using the
box norm inverse result. 
From the above argument, it is natural to try to prove that if ${\sf swap}(f)\geq \eps$, then there is a set $I\subseteq[n]$ and functions $g_1, h_1$ as above such that ${\sf swap}(f-\lambda_1 g_1h_1)$ is noticeably smaller than ${\sf swap}(f)$. Continuing in this fashion, one can hope to find sets $I_i\subseteq [n]$, functions $(g_i,h_i)$ 
and $\lambda_i\in (0,1]$ such that 
${\sf swap}(f-\sum\limits_{i=1}^{T}\lambda_i g_ih_i)$
is small, say smaller than $\eps/10$, and $T = O_{\eps}(1)$. In that case the triangle inequality implies
${\sf swap}(\sum\limits_{i=1}^{T}\lambda_i g_ih_i)\geq 9\eps/10$. 

It is useful to consider the simple case that $T=1$.
In this case, the last inequality means that ${\sf swap}(g_1){\sf swap}(h_1)={\sf swap}(g_1h_1)\geq 9\eps/10$, so either 
${\sf swap}(g_1)$ or ${\sf swap}(h_1)$ is significantly bigger than $\eps$. Since both functions can be viewed as random 
restrictions of $f$, we conclude that a suitably chosen random restriction of 
$f$ has swap-norm noticeably bigger than that of $f$. This facilitates an inductive approach.

In the case that $T>1$ the argument is not as straight-forward. Instead of showing that there is $i$ such that either $g_i$ or $h_i$ have noticeably bigger swap norms, one shows that there is complex combination 
$g'=\sum\limits_{i=1}^{T}\theta_i g_i$ or
$h'=\sum\limits_{i=1}^{T}\theta_i h_i$ 
with $\sum\limits_{i=1}^{T}\card{\theta_i}^2=1$ that
achieves swap-norm noticeably bigger than $\eps$. 
Additionally, one shows that if either $g'$ or $h'$ is correlated with a product function, then a random 
restriction of $f$ outside $I_1\cap\ldots\cap I_T$ 
is also correlated with a product function. Together,
these two facts facilitate an inductive approach as well.

We remark that the formal proof in~\cite{CSPs6} involves several significant technical challenges:
\vspace{-1ex}
\begin{enumerate}
    \item First, we do not
actually know how to show that the swap norm of 
$f-g_1h_1$ is noticeably smaller than the swap norm 
of $f$ itself, so it is not clear how to argue that 
the process will terminate within $T = O_{\eps}(1)$
steps based on this consideration alone. To address this issue one incorporates
an $L_2$-norm based energy increment approach, which
is useful since  
$\norm{f-g_1h_1}_2^2$ is indeed noticeably smaller than 
$\norm{f}_2^2$.\footnote{Combining the swap-norm and the 
$L_2$-norm energy increment strategies amount to 
defining a certain potential function that takes 
both into account, and measuring how it changes throughout the process.} 
\item Second, in the case $T>1$ 
obtaining a relation between $f$ and product functions
and $g'$, $h'$ is not clear for arbitrary functions 
$g_1,\ldots,g_T$ and $h_1,\ldots,h_T$. Instead, one uses SVDs of $f$ to find the components $g_i,h_i$. The upshot here is that taking $(g_1,h_1)$
to be SVD components of $f$ with respect to the partition $[n] = I_1\cup\bar{I}_1$, one has $g_1(x) = \E_{y\in_R\Sigma^{\bar{I}_1}}[f(x,y)\overline{h_1(y)}]$, 
so the correlation of $g_1$ with a product function $P$ can be related to the correlation of $f_{\bar{I}_1\rightarrow y}$ with $P$ for a random $y\in_R\Sigma^{\bar{I}_1}$. 
\item Third, 
the fact that the sets $I_i\subseteq [n]$ are distinct complicates the 
above argument considerably, as the functions $g_ih_i$
are not compatible with each other and it is not clear how to relate them to a common SVD of $f$. To address
this issue one applies a random restriction to ``project'' the sets $I_i$ into a common partition $(S,T)$, so that they (morally) all become SVD components with respect to the same partition.
\end{enumerate}
 
\section{Applications.}\label{sec:applications}
In this section we discuss some applications of the
results from~\Cref{sec:statements}.
\subsection{Additive Combinatorics.} 
We begin with applications in additive combinatorics, 
and more specifically to the study of restricted versions of the arithmetic progression patterns from~\Cref{sec:intro}.
\subsubsection{Restricted Arithmetic Progressions of Length $3$.}
\begin{definition}
    Let $p\geq 3$ be a prime and let $n\in\mathbb{N}$ be a large integer.
    \begin{enumerate}
        \item We say that a triplet $x,x+a,x+2a$
        forms a somewhat-restricted $3$-AP if 
        $x\in\mathbb{F}_p^n$ and $a\in \{0,1,2\}^n\setminus\{\vec{0}\}$. 
        We say that $A\subseteq \mathbb{F}_p^n$
        is somewhat-restricted $3$-AP free if 
        $A$ contains no somewhat-restricted $3$-AP.
        \item We say that a triplet $x,x+a,x+2a$
        forms a restricted $3$-AP if 
        $x\in\mathbb{F}_p^n$ and $a\in \{0,1\}^n\setminus\{\vec{0}\}$.
        We say that $A\subseteq \mathbb{F}_p^n$
        is restricted $3$-AP free if 
        $A$ contains no restricted $3$-AP.
    \end{enumerate}
\end{definition}
It is clear that for any set $A\subseteq \mathbb{F}_p^n$ we have that
\[
A\text{ is $3$-AP free}
~\Longrightarrow~
A\text{ is somewhat-restricted $3$-AP free}
~\Longrightarrow~
A\text{ is restricted $3$-AP free}.
\]
It is thus natural to ask: 
What is the largest density of a somewhat-restricted $3$-AP free set? What is the largest density of a restricted $3$-AP free set? It also makes sense to consider the counting versions of these problems: 
if $A$ has density $\alpha>0$ (thought of as a constant),
must it contain $\Omega_{\alpha}(1)$ 
fraction of the somewhat-restricted $3$-APs? Must it contain $\Omega_{\alpha}(1)$ 
fraction of the restricted $3$-APs?

One motivation
for studying these problems is that the Fourier-analytic approach, as outlined in~\Cref{sec:intro}, 
no longer works, so making progress on these problems is
likely to require a different approach. The density Hales-Jewett theorem~\cite{FKDHJ3,FKDHJ,PolyDHJ,SimpleDHJ} implies that a set $A\subseteq\mathbb{F}_p^n$ with no restricted $3$-APs has vanishing density $o(1)$. The quantitative bounds it gives though are either ineffective (if one uses the ergodic-theoretic proof of~\cite{FKDHJ3,FKDHJ}), or else very weak inverse-tower type (if one uses the combinatorial proof of~\cite{PolyDHJ,SimpleDHJ}). 
The counting versions, for which the density Hales-Jewett does not apply, have been raised in~\cite{HazlaHolensteinMossel2018} 
and in~\cite{Green2018_100OpenProblems} and remain
completely open.

In~\cite{BhangaleKhotMinzer2024} the authors 
combine~\Cref{thm:csps4_global} with an appropriate density increment argument and show: 
\begin{theorem}\label{thm:somewhat_restricted}
A somewhat-restricted $3$-AP free set $A\subseteq \mathbb{F}_p^n$ has density at most 
$O\left(\frac{1}{(\log\log\log n)^{\Omega_p(1)}}\right)$.
\end{theorem}

The proof of~\Cref{thm:somewhat_restricted} proceeds by first translating the assumption that $A$ is somewhat-restricted $3$-AP set to an inequality as~\eqref{eq:3AP}, except that $a$ is sampled uniformly from $\{0,1,2\}^n$. Using the fact that 
the uniform distribution over $\{(x,x+a,x+2a)~|~x\in\mathbb{F}_p,a\in\{0,1,2\}\}$ 
has no $(\mathbb{Z},+)$-embeddings they apply~\Cref{thm:csps4_global} to conclude that $f_A = 1_A - \mu(A)$ is correlated with a function of the form 
$L\cdot \chi\circ \sigma^{\otimes n}$. Using random restrictions one can get rid of the low-degree term $L$, and for simplicity we ignore it. 
As each $\chi_j$ is it a character over a group $G$ of size $O_p(1)$ there are only $W=O_p(1)$ many options 
for each $\chi_j$. Therefore, we may partition the coordinates $j\in [n]$ into groups $J_1,\ldots,J_{W}$ depending on $\chi_j$. The key observation is now that by taking $\card{G}$ coordinates in some $J_i$ and 
``identifying them'', namely plugging the same value of $v\in\mathbb{F}_p$ in all of them, the
function $\chi\circ \sigma$ no longer depends on them 
(as together they contribute $(\chi\circ \sigma(v))^{\card{G}} = 1$, where $\chi$ is the character applied on coordinates from $J_i$). 
Using this idea one can decompose $\mathbb{F}_p^n$
into copies of $\mathbb{F}_p^{n'}$ on which  $\chi\circ\sigma$ is constant and $n'\geq \Omega_p(n)$, giving a density increment for $A$ on one of these copies.

The case of restricted $3$-APs is more difficult, since the uniform distribution over
$\{(x,x+a,x+2a)~|~x\in\mathbb{F}_p,a\in\{0,1\}\}$
does contain $(\mathbb{Z},+)$-embeddings. Still, it
is pairwise-connected, and in particular one can apply~\Cref{thm:csps6} or~\Cref{thm:csps6_global}. 
Indeed, using these results along with a similar-in-spirit density increment argument,~\Cref{thm:somewhat_restricted} is strengthened in~\cite{CSPs6} to the following result:\footnote{We remark that similar results hold for a more general family of patterns of length $3$.}
\begin{theorem}\label{thm:restricted}
A restricted $3$-AP free set $A\subseteq \mathbb{F}_p^n$ has density at most 
$O\left(\frac{1}{(\log\log\log n)^{\Omega_p(1)}}\right)$.
\end{theorem}

\subsubsection{Combinatorial Lines of Length $3$.}
There is a variant of the $3$-APs pattern that is
even more restrictive than restricted $3$-APs. Due
to the lack of arithmetic structure, it is more
often referred to as \emph{combinatorial lines}.
\begin{definition}\label{def:comb_line}
    Let $k\geq 3$ be an integer. A combinatorial line
    of length $k$ in $\{0,1\ldots,k-1\}^n$ is a $k$-tuple $x_1,\ldots,x_k$ of distinct points in $\{0,1\ldots,k-1\}^n$ such that for each coordinate $j\in[n]$
    \[
    (x_1(j),\ldots,x_k(j))
    \in \{(0,\ldots,0), (1,\ldots,1), \ldots,(k-1,\ldots,k-1),(0,1,2,\ldots,k-1)\}.
    \]
\end{definition}
In words, a combinatorial line of length $k$ is a $k$-tuple $x_1,\ldots,x_k$ of distinct points such that on each coordinate $j$, either all entries are equal, or else
they must satisfy $x_1(j)=0,x_2(j) = 1,\ldots,x_k(j) = k-1$.

The
problem of determining the density of the largest $A\subseteq\{0,1,\ldots,k-1\}^n$ with no combinatorial lines of length $k$ is known in the literature as the density Hales-Jewett problem, or DHJ-$k$ in short (being the density version of the classical Hales-Jewett theorem~\cite{HalesJewett1963}). 
It has received significant attention over the years 
for several reasons. First, it is arguably the most strict combinatorial structure one may hope to find inside any dense set $A$. Second, one can show that 
a set of integers $A\subseteq\{1,\ldots,n\}$ with no
$k$-APs can be translated into a set of points $A'\subseteq\{0,1\ldots,k-1\}^{n'}$ with similar density and no combinatorial lines of length $k$. 
In particular, showing that a set with no combinatorial lines of length $k$ must have vanishing density implies Szemer\'edi's theorem.

Using ergodic-theoretic techniques, Furstenberg and Katznelson showed that
a set $A\subseteq\{0,1\ldots,k-1\}^n$
with no combinatorial lines of length 
$k$ has vanishing density. The first work to establish quantitative bounds was done by the Polymath1 project~\cite{PolyDHJ}, who established inverse-tower type density bounds. For the case $k=3$, their bound was of the order $1/\sqrt{\log^{*} n}$, where $\log^* n$ is the number of times that $\log$ needs to be applied on $n$ until one gets below $1$.
In~\cite{DHJ3} the authors establish the first known ``reasonable bound'' for the DHJ-$3$ problem, meaning a bound with finitely many logarithms:
\begin{theorem}\label{thm:DHJ3}
A set $A\subseteq \{0,1,2\}^n$ with no combinatorial lines of length $3$ has density at most 
$O\left(\frac{1}{(\log\log\log\log n)^{\Omega(1)}}\right)$.
\end{theorem}

The proof of~\Cref{thm:DHJ3} uses~\Cref{thm:csps6} in non-obvious way; in fact, 
the proof also uses~\Cref{thm:csps7}. The main issue
is that distributions $\mu$ relevant to the study
of combinatorial lines are only supported
on $\{(0,0,0), (1,1,1), (2,2,2), (0,1,2)\}$, so they are not pairwise-connected. Therefore, one cannot (directly) apply~\Cref{thm:csps6} to get any structural result on $f_A = 1_A - \card{A}/3^n$ if $A$ has no combinatorial lines of length $3$. To remedy this, the authors use ideas
from Shkredov's argument for the corners problem~\cite{Shkredov2006Corners} 
(see also~\cite{Green2005FiniteFieldModels,Green2005ShkredovCorners}). In the corners problem one wants to upper bound
the density of a set $A\subseteq\mathbb{F}_p^n\times\mathbb{F}_p^n$ 
that contains no pattern of the form $(x,y), (x+a,y), (x,y+a)$. For the integer version of the problem, a direct density increment approach was used by~\cite{AjtaiSze}, who showed that the density of a corner-free set is $o(1)$. The quantitative bound was 
significantly improved by Shkredov's~\cite{Shkredov2006Corners} via a \emph{relative} density increment argument. His idea
was that, instead of measuring the density of $A$ in comparison to the ambient space, one should measure it with respect to a rectangle $E_1\times E_2$ where $E_1,E_2\subseteq\mathbb{F}_p^n$. The upshot of this approach is that it is relatively straightforward to show that if $A\subseteq \mathbb{F}_p^n\times \mathbb{F}_p^n$ has no corners, then it admits a density increment on some $E_1\times E_2$ for sizable $E_1,E_2$. The downside is that it is tricky to iterate this sort of density increment statement. To 
do so, Shkredov ensures that $E_1,E_2$ satisfy suitable pseudo-randomness properties, which then allows him to iterate the argument (see~\cite{JaberLiuLovettOstuniSawhney2025} for a recent improvement). 

The proof of~\Cref{thm:DHJ3} follows the high-level
approach of Shkredov's argument. The argument uses
a ``dictionary'' between notions in the context of corners (such as rectangles) to notions in the context of DHJ (such as $0,1$ and $0,2$-insensitive sets and the disjoint product), that was in fact already observed by~\cite{PolyDHJ}. 
The main high-level difference between the argument of~\cite{DHJ3} and of~\cite{Shkredov2006Corners} is in the pseudo-random notion that is used. In the context of corners, the correct notion turns out to be having small $U^2$ uniformity norm, whereas in the context of DHJ-$3$, the correct notion turns out to (morally) be having a small swap norm.\footnote{Strictly speaking, for technical reasons the necessary pseudo-randomness condition is that the function has no correlation with product functions
as in~\Cref{def:prod_fn}, even after randomly restricting all but $\sqrt{n}$ of the coordinates.}

\subsection{Complexity Theory.} 
We next discuss some applications to problems in complexity theory.
\subsubsection{The Hybrid Algorithm for Satisfiable CSPs.}
In~\Cref{thm:raghavendra}, Raghavendra shows a dichotomy result for approximating CSPs with a small loss in the completeness parameter. 
As explained therein, a key component in his proof is 
a dictatorship test  with matching parameters. 
In~\cite{CSPs7} the authors show that Raghavendra's 
dictatorship test construction remains valid with no
loss in the completeness parameter for predicates $P\colon \Sigma^k\to\{0,1\}$ such that $P^{-1}(1)$
has no Abelian embeddings:
\begin{theorem}\label{thm:predicates_noembed}
    Suppose that $\mathcal{P}$ is a collection of 
    $k$-ary predicates such that $P^{-1}(1)$ has no Abelian embeddings for all $P\in\mathcal{P}$.\footnote{Strictly speaking, the required condition is more technical and asserts that in an optimal SDP solution, all local distributions have no Abelian embeddings. We refrain from stating it explicitly for the sake of simplicity.} Then there exists $s\in [0,1]$ such that:
    \vspace{-1ex}
    \begin{enumerate}
        \item Algorithm: there is a polynomial time algorithm that solves gap-$\mathcal{P}$-CSP$[1,s]$.
        \item Evidence for Hardness: there is a dictatorship test for $\mathcal{P}$ with completeness $1$ and soundness $s+o(1)$.
    \end{enumerate}
\end{theorem}

The algorithm in~\Cref{thm:predicates_noembed} is the same as in~\Cref{thm:raghavendra} and is based on semi-definite programming.
For predicates that have Abelian embeddings, 
it is well known that the optimal approximation algorithm cannot be based only on semi-definite programming relaxations~\cite{Grigoriev2001Parity,Schoenebeck2008Lasserre}. As of the time of writing, there is no clear candidate approximation algorithm in this case, and many special cases are still open. 

In~\cite{CSPs5} the authors consider $3$-ary predicates $P$ that admit Abelian embeddings but 
no $(\mathbb{Z},+)$-embeddings. For such predicates 
$P$ the authors propose a polynomial time ``hybrid algorithm'', which is an intertwined combination of a semi-definite programming algorithm and a Gaussian elimination algorithm, as a candidate optimal approximation algorithm. Their result reads:
\begin{theorem}\label{thm:csps5}
    Suppose that a collection of $3$-ary predicates $\mathcal{P}$ satisfies that $P^{-1}(1)$ has no $(\mathbb{Z},+)$-embeddings for all $P\in\mathcal{P}$.\footnote{Strictly speaking, the required condition is more technical and asserts that in an optimal SDP solution, all local distributions have no $(\mathbb{Z},+)$-embeddings. We refrain from stating it explicitly for the sake of simplicity.} Then there exists $s\in [0,1]$ such that:
    \vspace{-1ex}
    \begin{enumerate}
        \item Algorithm: there is a polynomial time algorithm for gap-$\mathcal{P}$-CSP$[1,s]$.
        \item Evidence for Hardness: there is a dictatorship test for $\mathcal{P}$ with completeness $1$ and soundness $s+o(1)$.
    \end{enumerate}
\end{theorem}
While we do not outline the proof of~\Cref{thm:csps5}, we mention that it uses a 
generalization of the invariance principle of~\cite{MOO} called the \emph{mixed invariance principle}. Whereas the invariance principle of~\cite{MOO} relates the behavior of low-degree polynomials between discrete space and Gaussian space, 
the mixed invariance principle relates the behavior of the wider 
family of ``low-degree polynomials times character 
function'' (as appearing in~\Cref{thm:csps4_global})
between discrete space and the product of Gaussian and Abelian group spaces.

\subsubsection{Multiplayer Parallel Repetition.}\label{sec:mpr}
The parallel repetition theorem of Raz~\cite{Raz} is a powerful tool in theoretical computer science. To demonstrate it, consider a $2$-player game based on graph coloring: suppose that $G = (V,E)$ is a graph that is either $3$-colorable (meaning there is $\chi\colon V\to\{0,1,2\}$ such that $\chi(u)\neq \chi(v)$ for all $(u,v)\in E$) or is $\eps$-far from being $3$-colorable (meaning that for all $\chi\colon V\to\{0,1,2\}$ it holds that $\chi(u)= \chi(v)$ for at least $\eps$ fraction of the edges $(u,v)\in E$). Consider the graph $G^{\otimes n}$ whose vertices are $V^n$, and whose edges are $((u_1,\ldots,u_n),(v_1,\ldots,v_n))$ such that $\forall i~(u_i,v_i)\in E$. Suppose we wish to label $V^n$ by $\{0,1,2\}^n$ so 
that for each edge $(\vec{u},\vec{v})$, the assignment of the endpoints differ on all coordinates.
If $G$ is $3$-colorable, it is easily seen that such assignment for $G^{\otimes n}$ exists. The parallel repetition theorem asserts that if $G$ is $\eps$-far from being $3$-colorable, then any assignment to $G^{\otimes n}$ satisfies at most $(1-\Omega_{\eps}(1))^{n}$ fraction of the edges.

Proving multiplayer parallel repetition theorems turns out to be significantly more challenging.
Here, a $k$-player game $\inst$ consists of a $k$-partite
$k$-uniform hypergraph $G = (V_1\cup\ldots\cup V_k,E)$, alphabets $\Sigma_1,\ldots,\Sigma_k$ and a constraint 
$\Phi_e \subseteq \Sigma_1\times\ldots\Sigma_k$ for each edge $e\in E$. The value of $\inst$ is defined as
\[
\val(\inst) = \max_{A_i\colon V_i\to \Sigma_i}\frac{\card{\{e=(v_1,\ldots,v_k)\in E~|~(A_1(v_1),\ldots,A_k(v_k))\in \Phi_e\}}}{\card{E}}.
\]
The $n$-fold repeated game $\inst^{\otimes n}$ 
is given by the $k$-partite $k$-uniform hypergraph $({V_1}^n\cup\ldots\cup {V_k}^n, E')$ with the edge set 
$E' = \{(\vec{v}_1,\ldots,\vec{v}_k)~|~(\vec{v}_1(j),\ldots,\vec{v}_k(j))\in E~\forall j=1,\ldots,n\}$, alphabets $\Sigma_1^n,\ldots,\Sigma_k^n$ and constraints $\Phi_{e'}\subseteq \Sigma_1^n\times\ldots\times \Sigma_k^n$
for $e'\in E'$. We view each a label from $\Sigma_1^n$ to a vertex 
$\vec{v}_1$ as assigning a label from $\Sigma_1$ to each one of the coordinates of $\vec{v}_1$. In this language, the constraint $\Phi_{e'}$ on an edge 
$e' = (\vec{v}_1,\ldots,\vec{v}_k)$ imposes that for each $j=1,\ldots,n$, the labels given to the $j$th coordinate of $\vec{v}_1,\ldots,\vec{v}_k$ satisfy the constraint $\Phi_{e_j}$ where 
$e_j = (\vec{v}_1(j),\ldots,\vec{v}_k(j))\in E$.

With this terminology,~\cite{FortnowRompelSipser1994} suggested to study the multiplayer parallel repetition conjecture: if
$\inst$ is a $k$-player game such that
${\sf val}(\inst)\leq 1-\eps$ and $\eps>0$, then 
${\sf val}(\inst^{\otimes n})\leq (1-\eps')^n$
where $\eps'>0$ depends only on $k,\eps$ (and possibly other parameters of the base game $\inst$ such as the alphabet sizes). In an important work Raz~\cite{Raz} showed, using information theoretic techniques, that this statement is correct for $k=2$. The case $k\geq 3$ remains wide open to date. The techniques of Raz work for the class of ``connected'' $k$-player games~\cite{DinurHarshaVenkatYuen2017}, but they fail for any disconnected game (the definition of a connected game is analogous to the definition of a connected distribution).

Recent works~\cite{GHZ,XORgames,ProjGames} made some progress on the multiplayer parallel repetition conjecture, and used the analytic machinery discussed in~\Cref{sec:statements} to prove
exponential decay rates for some classes of $3$-player games.

\section{Open Problems.}\label{sec:open}
We finish this article by mentioning a few open
problems for future research. The first problem 
is to prove a general inverse theorem for correlations over pairwise-connected distributions:
\begin{problem}\label{prob:inv}
    Resolve~\Cref{conj:general_inverse}.
\end{problem}
We remark that even resolving special cases of~\Cref{conj:general_inverse} (for $k\geq 4$)
would be significant progress. The next problem 
is to establish effective bounds for the density Hales-Jewett problem for $k\geq 4$:
\begin{problem}\label{prob:dhj}
    Show that if $A\subseteq\{0,1\ldots,k-1\}^n$ 
    has no combinatorial lines, then the density of $A$ is at most $O\left(\frac{1}{\log\ldots\log n}\right)$ where the number of applications of $\log$ is $O_k(1)$.
\end{problem}
We expect that to resolve~\Cref{prob:dhj} one would (at the very least) need to resolve~\Cref{prob:inv}, 
but suspect more work to be needed. We believe it would be already interesting to find a density increment strategy (or any other strategy) that resolves~\Cref{prob:dhj} assuming a statement along the lines of~\Cref{conj:general_inverse}.

Many developments in the study of the Gowers uniformity norms have been motivated by ergodic theory, and we believe that there may be an interesting ergodic theoretic view of the general correlations studied in this article. It would be interesting to find such a
connection and perhaps use it to prove inverse type results in the spirit of~\Cref{thm:csps4,thm:csps6} 
(even with no quantitative bounds):
\begin{problem}\label{prob:ergodic}
    Find an ergodic-theoretic approach for the study of $k$-wise correlations as in~\eqref{eq:main_kwise_cor} and for establishing inverse theorems as discussed in this article.
\end{problem}

From the perspective of theoretical computer science, 
perhaps the most ambitious goal with regards to the topics in this article would be to establish a dichotomy result for approximating satisfiable CSPs:
\begin{problem}\label{prob:sat_CSP}
    Show that the following dichotomy result: for all $k\in\mathbb{N}$, finite alphabet $\Sigma$ and $\mathcal{P}$ collection of $k$-ary predicates over $\Sigma$ such that $\mathcal{P}$-CSP is NP-hard, there is an $s\in(0,1)$ such that:
    \begin{enumerate}
        \item Algorithm: there is a polynomial time algorithm for gap-$\mathcal{P}$-CSP$[1,s]$.
        \item Hardness: for all $\delta$, the problem gap-$\mathcal{P}$-CSP$[1,s+\delta]$ is NP-hard, possibly assuming a conjecture such as the Rich $2$-to-$1$ Games Conjecture~\cite{BKM}. A bit less ambitiously, show a dictatorship test for $\mathcal{P}$ with completeness $1$ and soundness $s+\delta$.
    \end{enumerate}
\end{problem}

Next, it would be interesting to find an analytical proof of the dichotomy theorem of Zhuk and Bulatov~\cite{Zhuk,Bulatov}. Besides being a different proof it may be the case that an analytical approach would give a different (possibly simpler) algorithm:
\begin{problem}\label{prob:dich}
    Give an analytical proof for the dichotomy theorem: for all $k\in\mathbb{N}$, finite alphabet $\Sigma$ and a collection of $k$-ary predicates $\mathcal{P}$ over $\Sigma$, the problem $\mathcal{P}$-CSP either admits a polynomial time algorithm or else is NP-hard.
\end{problem}
We expect both~\Cref{prob:sat_CSP} and~\Cref{prob:dich} to be very challenging, and
think that analyzing specific classes of predicates 
would already be very interesting. In particular, at
present it is not clear what an algorithm as in~\Cref{prob:sat_CSP} should be.

The next problem is to make further progress on the multiplayer parallel repetition conjecture. The results discussed in~\Cref{sec:mpr} apply only to 
the class of XOR games, and it is not clear how 
to use any of these analytical techniques for other classes of multiplayer games (even for a small number of players, say $3$):
\begin{problem}\label{prob:mpr}
    Show that if $\inst$ is a $k$-player game with 
    $\val(\inst)\leq 1-\eps$, then $\val(\inst^{\otimes n})\leq (1-\eps')^n$ where $\eps' = \eps'(\eps,\inst,k)>0$.
\end{problem}
The best known result in the direction of~\Cref{prob:mpr} is by~\cite{Verbitsky1996PRC}, who showed that $\lim_{n\rightarrow\infty}\val(\inst^{\otimes n})=0$, with no explicit quantitative bounds. 
Establishing any reasonable bounds in~\Cref{prob:mpr}
would already be significant.

\paragraph{Tensorization in different contexts?} 
It is a very common phenomenon that tensorizing inequalities that are not captured by eigenvalues of matrices is 
very challenging. In fact, this is partly the reason  problems in multiplayer communication complexity, and regarding explicit constructions of high-rank tensors, are so challenging.
We believe it would be very interesting to see if there are any approaches along the lines described in~\Cref{sec:proofs} that may be applicable in 
these contexts (or others). 
By that, we mean an approach that 
finds a non-trivial base case statement which is 
more amendable to tensorization, and then tensorizes
it a non-automatic way.
\bibliographystyle{siamplain}
\bibliography{ref}
\end{document}